\documentclass[11pt,a4paper,english]{article}
\usepackage{jheppub}
\usepackage[T1]{fontenc}
\usepackage[latin9]{inputenc}
\usepackage{esint}
\usepackage{babel}

\usepackage{pstricks}
\usepackage{bbm}

\definecolor{mypurple}{rgb}{1,0,1}

\def\Re{\mathop{\rm Re}}
\def\Im{\mathop{\rm Im}}
\def\ignore#1{{}}
\def\vec#1{\mbox {\boldmath $#1$}}
\def\svec#1{\mbox {\scriptsize\boldmath $#1$}}
\def\floor#1{\lfloor #1\rfloor}
\def\dn{\mathop{\rm dn}}
\def\sech{\mathop{\rm sech}}

\providecommand{\tabularnewline}{\\}

\title{Finite-Volume Spectra of the Lee-Yang Model}

\author[a]{Zoltan Bajnok}
\author[a,b]{Omar el Deeb}
\author[c]{and Paul A. Pearce}

\affiliation[a]{MTA Lend\"ulet Holographic QFT Group, Wigner Research Centre for Physics\\ 
H-1525 Budapest 114, P.O.B. 49, Hungary}
\affiliation[b]{Physics Department,  Faculty of Science,  Beirut Arab University (BAU)\\
Beirut, Lebanon}
\affiliation[c]{Department of Mathematics and Statistics, University of Melbourne\\
Parkville, Victoria 3010, Australia}

\emailAdd{bajnok.zoltan@wigner.mta.hu}
\emailAdd{o.deeb@bau.edu.lb}
\emailAdd{P.Pearce@ms.unimelb.edu.au}

\abstract{We consider the non-unitary Lee-Yang minimal model ${\cal M}(2,5)$
in three different finite geometries: {(i)}~on the interval with integrable
boundary conditions labelled by the Kac labels $(r,s)=(1,1),(1,2)$,
{(ii)}~on the circle with periodic boundary {conditions} and {(iii)}~on the periodic circle including
an integrable purely transmitting defect. We apply $\varphi_{1,3}$
integrable perturbations on the boundary and on the defect and describe
the flow of the spectrum. {Adding a} $\Phi_{1,3}$
integrable perturbation {to move off-criticality} in the bulk, we determine
the finite size spectrum of the massive scattering theory in the three
geometries via {Thermodynamic Bethe Ansatz} (TBA) equations. We derive
these {integral} equations for all excitations by solving, in the continuum scaling
limit, the TBA functional equation{s} satisfied by the transfer matrices
of the associated $A_{4}$ {RSOS} lattice model of Forrester and Baxter in
Regime~III. The excitations are classified in terms of $(m,n)$
systems. The excited state TBA equations agree with the previously
{conjectured} equations in the boundary and periodic cases. 
{In the defect case, 
new TBA equations confirm} previously
conjectured transmission factors.}


\arxivnumber{2014.xxxx}

\begin{document}

\maketitle
\flushbottom

\section{Introduction}

The determination of the full spectrum of a 1+1 dimensional Quantum
Field Theory (QFT) in finite volume is a highly non-trivial and usually
{intractable} task. Even the vacuum energy has a complicated volume dependence,
which generally cannot be calculated exactly.{}
For integrable theories the situation is better{.} The bootstrap approach
provides exact expressions for the masses of the particles together
with their scattering matrices. These infinite volume quantities 
can be used to determine an approximate spectrum via Bethe-Yang {(BY)} equations~\cite{BY,YangYang}
for large volumes. The BY finite size spectrum contains all polynomial
correction{s} in the inverse {powers} of the volume but neglects exponentially
small vacuum polarization effects. The vacuum polarization effects
can be expressed in terms of the scattering matrix and{,} for the ground-state{,}
they can be calculated exactly by the {T}hermodynamic Bethe Ansatz (TBA) method. 
The TBA method exploits the fact that for{,} large Euclidean time{,} the partition
function is dominated by the contribution of the ground state. Exchanging
the role of space and Euclidean time{,} the partition function {only needs} to
be evaluated in the large volume limit, where Bethe-Yang equations
are {accurate}. Calculating the partition function in the saddle point
approximation{,} integral equations (TBA) can be derived for the saddle
point particle densities (pseudo-energies). The solutions of these
nonlinear TBA equations provides the ground-state energy \cite{Zam90,Zam91}. 

{The TBA method of exploiting the invariance properties of the partition function does not easily extend to} 
excited states. Nevertheless, the exact ground-state TBA equations
can be used to gain information about {certain} excited states \cite{Dorey} 
{under the assumption that these} excited states and the ground-state are related
by {analytic continuation in a suitable variable}. Carefully analysing the {analytic} behaviour
of the TBA equations for complex volumes {for these special excited states,} TBA equations for zero momentum 
two particle states {are} obtained. The {key} difference{,} compared to the
ground-state equation{,} is in the {appearance of} so called source terms, or equivalently,
in choosing a different contour for integrations. {Such a program using analytic continuation has not been successfully carried out to obtain} 
TBA equations for the full {excitation spectrum even for the simple non-unitary scaling Lee-Yang model~\cite{LeeYang}.} 
Although, from the explicitly calculated cases, a natural conjecture for all exited states can be formulated. The Lee-Yang 
theory describes~\cite{Fisher78,Cardy85,CardyMuss89} the closing, in the complex magnetic field plane, of the gap 
in the distribution of Lee-Yang zeros of the two-dimensional Ising model. 

There is a {more general and systematic way to obtain} TBA integral equations for
excited states based on {functional relations~\cite{PearK91,KlumP91,KlumP92} coming from Yang-Baxter integrable~\cite{BaxBook} lattice regularizations. These functional equations take the form of fusion, $T$- and $Y$-sytems}. 
{The $Y$-system involves the pseudo-energies and, at criticality, it describes the conformal spectra. It is {\it universal\/} in 
the sense that the same equations hold for all geometries and all excitations.}
{Relevant physical solutions for the excitations are selected out by applying different 
asymptotic and analyticity properties to} 
the $Y$-functions. {Indeed,} knowing {this asymptotic and analytic information,} the functional
relations can be {recast into TBA integral} equations for the full {excitation} 
spectrum. {This program has been successfully carried out to completion~\cite{OPW97} for the tricritical Ising model ${\cal M}(4,5)$ with conformal boundary conditions.}
The lattice regularization approach is not limited to CFT but also extends to  
integrable QFTs. In \cite{PearN98} the ground-state TBA equations
of the periodic $A$ and $D$ RSOS models were derived, while in \cite{PCAI,PCAII}
the full spectrum of the tricritical Ising model was described on
the interval. The integral equations for the spectrum of the sine-Gordon
theory underwent a parallel development. The ground-state equation was derived
in \cite{DdV} and extended to some excited states in \cite{DdV2,FMQR}.

The functional {form of the $Y$-system reflects} the integrable structure of the  Conformal Field Theory (CFT). 
{In principle, the $Y$-system can be derived~\cite{BLZ} directly in the continuum scaling limit}. 
{Within the lattice approach, it is obtained, in a more pedestrian way, by taking the continuum scaling limit 
of an integrable lattice regularization of the theory}. 
{\ignore{Defining the model as a continuum limit of a lattice model the functional
relations appear from the fusion hierarchy of transfer matrices, which
defines the lattice spectrum.}}{A distinct advantage of the lattice approach 
is that it explicitly provides the asymptotic and analytic} 
properties of the $Y$-functions, 
{which otherwise need to be guessed.} 
{More specifically}, the lattice {approach provides the relevant asymptotic and analyticity properties and hence the complete classification of all excited states} of the theory. 

The Lee-Yang model ${\cal M}(2,5)$, 
perturbed by its only relevant spinless perturbation, is perhaps the
simplest theory of a single massive particle and so is usually the
first model studied to understand properties of massive theories.
The scattering matrix is merely a simple CDD factor with a self-fusing
pole
\begin{equation}
S(\theta)=\frac{\sinh\theta+i\sin\frac{\pi}{3}}{\sinh\theta-i\sin\frac{\pi}{3}}\label{Smatrix}
\end{equation}
The TBA equation for the model was derived by Zamolodchikov \cite{Zam90,Zam91}
\begin{equation}
\epsilon(\theta)=mL\cosh\theta+s(\theta)-\int_{-\infty}^{\infty}\frac{d\theta^{'}}
{2\pi}\,\varphi(\theta-\theta^{'})\log(1+e^{-\epsilon(\theta^{'})})
\end{equation}
where the kernel is related to the scattering matrix by $\varphi(\theta)=-i\partial_{\theta}\log S(\theta)$. 
For the ground state energy
\begin{equation}
E(L)=e(\theta)-m\int_{-\infty}^{\infty}\frac{d\theta}{2\pi}\,\cosh\theta\,
\log(1+e^{-\epsilon(\theta)})
\end{equation}
with vanishing source terms $s(\theta)=0$ and $e(\theta)=0$. Later,
numerical investigation of the analytically continued TBA
ground-state solution, for complex volumes, led \cite{Dorey}
to TBA equations for certain excited states with source terms
\begin{equation}
s(\theta)=\sum_{i}\log\frac{S(\theta-\theta_{i})}{S(\theta-\theta_{i}^{*})}\label{eq:tbapersource}
\end{equation}
whose parameters are determined by the equations 
\begin{equation}
\epsilon(\theta_{j})=i(2n_{j}+1)\pi
\end{equation}
The contribution to the energy is $e(\theta)=-im\sum_{i}(\sinh\theta_{i}-\sinh\theta_{i}^{*})$. 
The $Y$-system \cite{Zam91}
\begin{equation}
Y(\theta)=e^{\epsilon(\theta)},\qquad Y(\theta+\frac{i\pi}{3})Y(\theta-\frac{i\pi}{3})=1+Y(\theta)
\end{equation}
 of the Lee-Yang model has been recast \cite{BLZ}
as an integral equation by assuming the analytic properties of the
$Y$-function thus providing a conjectured exact finite volume spectrum
for periodic boundary conditions. 

The Lee-Yang model has also been used as a prototypical example to
extend integrability into other space-time geometries. The ground-state
energy of the Lee-Yang model on the interval was derived in \cite{LMSS}
which, additionally to the $L\to2L$ and $E\to2E$ replacements, contained
the left/right reflection factors $R_{\alpha/\beta}(\theta)$ in the
source term 
\begin{equation}
s(\theta)=-\log\Big[ R_{\alpha}(\frac{i\pi}{2}+\theta)R_{\beta}(\frac{i\pi}{2}-\theta)\Big]
\end{equation}
The analytic continuation method provided excited state TBA equations
in \cite{DPTW,DRTW} with additional sources of the
form (\ref{eq:tbapersource}), in which each $\theta_{i}$ is paired with a partner 
$-\theta_{i}$. Integrability also extends to include integrable
defects of the Lee-Yang model. Indeed, the ground-state defect TBA
equations were derived in \cite{BS} with the source
term being the logarithm of the transmission factor:
\begin{equation}
s(\theta)=-\log T_{+}(\frac{i\pi}{2}-\theta)
\end{equation}

Perhaps surprisingly, the lattice regularization approach has not been systematically 
developed for the simple Lee-Yang theory and other non-unitary minimal models~\cite{BPZ84}. 
Our aim is to fill this gap for the Lee-Yang theory. 
In the present paper we
give a complete analysis of the finite volume spectrum of the Lee-Yang
model in the various geometries using a lattice approach.  
{It is expected that the methods developed here will apply to other non-unitary minimal models.}

The paper is organized as follows: 
In Section 2, we introduce the Lee-Yang theory
as the continuum scaling limit~\cite{Huse84,Riggs89} of the $A_{4}$ RSOS lattice model of Forrester-Baxter~\cite{ABF84,FB85}
in Regime~III with crossing parameter $\lambda=\frac{3\pi}{5}$.
We set up commuting transfer matrices (i)~with 
periodic boundary conditions, (ii)~with an integrable defect (called a
seam in the lattice terminology) and (iii)~with two integrable boundaries.
By properly normalizing the transfer matrices we show that they all
satisfy the same universal functional relation in the form of a $Y$-system. 
The conformal spectra of these transfer matrices are analysed in Section 3. We  
investigate the analytic structure of the transfer matrix eigenvalues and classify 
all excited states. We do this first for the trigonometric theory whose scaling
limit corresponds to the conformal Lee-Yang model. For the three geometries, 
we give detailed correspondences between paths, Virasoro descendants and 
patterns of zeros. The seam and boundaries admit a field which corresponds 
to the $\varphi_{1,3}$ perturbation.
Appropriately varying this field implements~\cite{FPR} renormalization group
flows connecting different conformal defect/boundary fixed points. 
We analyse the flow of the spectrum using the language of the paths, Virasoro 
descendants and zeros of the transfer matrix eigenvalues. 
Changing the trigonometric 
Boltzmann weights to suitable elliptic weights induces a massive
$\Phi_{1,3}$ perturbation. We round out this section by also classifying the states 
of the massive theory. In Section 4, we combine the analytic information with the
functional relations to derive integral TBA equations for the full finite
volume spectrum in the various geometries in the critical case. 
The off-critical TBA equations are derived in Section~5.  
Finally, we conclude with some discussion in 
Section 6.

\section{Lee-Yang Lattice Model and Transfer Matrices}

The Lee-Yang lattice model is a Restricted Solid-On-Solid (RSOS) model~\cite{ABF84,FB85} 
defined on a square lattice with heights $a=1,2,3,4$ restricted so that nearest neighbour heights
differ by $\pm1$. The heights thus live on the $A_{4}$ Dynkin diagram.
It is convenient to regard the Lee-Yang model as a special case $L=4$ of the more general 
$A_{L}$ Forrester-Baxter~\cite{FB85} models.

\subsection{Lee-Yang lattice model as the $A_{4}$ {RSOS} model}

The Boltzmann weights of the $A_{L}$ Forrester-Baxter~\cite{FB85} models with heights $a=1,\ldots,L$
are
\begin{eqnarray}
W\Big(\Big.\begin{array}{cc}
a\pm1 & a\nonumber\\
a & a\mp1
\end{array}\Big|u\Big) & = & \frac{s(\lambda-u)}{s(\lambda)}\\
W\Big(\Big.\begin{array}{cc}
a & a\pm1\label{RSOS}\\
a\mp1 & a
\end{array}\Big|u\Big) & = & \frac{g_{a\mp1}}{g_{a\pm1}}\Big(\frac{s((a\pm1)
\lambda)}{s(a\lambda)}\Big)\,\frac{s(u)}{s(\lambda)}\\
W\Big(\Big.\begin{array}{cc}
a & a\pm1\nonumber\\
a\pm1 & a
\end{array}\Big|u\Big) & = & \frac{s(a\lambda\pm u)}{s(a\lambda)}.
\end{eqnarray}
 Here $s(u)=\vartheta_{1}(u,q)$ is a standard elliptic theta function~\cite{GR}

\begin{eqnarray}
\vartheta_{1}(u,q)=2q^{1/4}\sin u\prod_{n=1}^{\infty}(1-2q^{2n}\cos2u+q^{4n})(1-q^{2n})
\end{eqnarray}
 where $u$ is the spectral parameter and the square of the elliptic nome $t=q^2$ is
a temperature-like variable corresponding to the $\Phi_{1,3}$ integrable
bulk perturbation. The crossing parameter $\lambda$ is 
\begin{equation}
\lambda=\frac{(m'-m)\pi}{m'}\label{crossing}
\end{equation}
 where $m<m'$, $m'=L+1$ and $m,m'$ are coprime integers. Integrability
derives from the fact that these local face weights satisfy the Yang-Baxter
equation. The gauge factors $g_{a}$ are arbitrary but we will take
$g_{a}=1$.
With this choice, the face weights are only symmetric under reflections
about the NW-SE diagonal. Notice also that, in the non-unitary
cases ($m'\ne m+1$), some Boltzmann weights are negative.
At criticality, $q=0$ and  the function $s(u)$
can be taken to be $s(u)=\sin u$. 

In the continuum scaling limit,
the critical Forrester-Baxter models in Regime~III
\begin{equation}
\mbox{Regime~III:}\qquad 0<u<\lambda,\qquad 0<q<1 
\end{equation}
are associated with the minimal
models ${\cal M}(m,m')$ with central charges 
\begin{equation}
c=1-\frac{6(m-m')^{2}}{mm'}
\end{equation}
In this paper, we only consider the Lee-Yang model 
\begin{equation}
{\cal M}(2,5):\qquad\lambda=\frac{3\pi}{5}
\end{equation}
with central charge $c=-22/5$. The associated conformal Virasoro algebra 
admits two irreducible representations
with conformal weights $h_{1,1}=h_{1,4}=0$, $h_{1,2}=h_{1,3}=-1/5$. 
It follows that the specific heat exponent $\alpha$ and correlation length exponent $\nu$ of the classical $d=2$ dimensional Lee-Yang model are given by the critical exponent scaling laws
\begin{equation}
h_{1,3}=\frac{1-\alpha}{2-\alpha}, \qquad 2-\alpha=d\nu=2\nu,\qquad \alpha=\frac{7}{6},\qquad \nu=\frac{5}{12}
\end{equation}

\subsection{Transfer matrices and functional relations}

In this section, we construct transfer matrices $T(u)$ from
the local face weights. Since the local face weights satisfy the Yang-Baxter
equations, the transfer matrices form commuting families $[T(u),T(v)]=0$
from which integrable one-dimensional quantum chains can be derived~\cite{BEPR} in the Hamiltonian limit. 
These transfer matrices satisfy the functional relations
\begin{equation}
\vec T(u)\vec T(u+\lambda)=\mathbbm{1}+\vec Y\vec T(u+3\lambda)\label{eq:funcrel}
\end{equation}
where $\vec Y$ is the $\mathbb{Z}_2$ height reversal matrix. 
Apart from the change in the value of the crossing parameter $\lambda$ from $\frac{\pi}{5}$ to $\frac{3\pi}{5}$, this is the same functional equation~\cite{BaxP82,BaxP83} that holds for the tricritical
hard square and hard hexagon models~\cite{Bax80,BaxBook}. This simple 
change in the crossing parameter, however, drastically changes the relevant
analyticity properties of the model. 

The conformal spectra $E_{n}$ of the Lee-Yang model in different geometries  
can be extracted from the logarithm of the transfer matrix eigenvalue 
via finite-size corrections~\cite{BCN}. For periodic systems (with or without a seam) or in the presence of a boundary, the finite-size corrections are given respectively by
\begin{align}
\mbox{}\hspace{-.2in}-\!\log T(u)&=Nf_{\text{\tiny bulk}}(u)+f_{\text{\tiny s}}(u,\xi)+\frac{2\pi}{N}\big[(E_n\!+\!\bar{E}_n)\sin\vartheta+i(E_n\!-\!\bar{E}_n)\cos\vartheta\big]+o\big(\frac{1}{N}\big)\\
\mbox{}\hspace{-.2in}-\!\log T(u)&=2Nf_{\text{\tiny bulk}}(u)+f_{\text{\tiny b}}(u,\xi)+\frac{2\pi}{N}\, 
        E_{n}\sin\vartheta+o\big(\frac{1}{N}\big)
\end{align}
where $N$ is the number of columns, $\vartheta=\tfrac{5u}{3}$ is the anisotropy angle, $f_{\text{\tiny bulk}}(u)$ is the bulk free energy per column and $f_{\text{\tiny s/b}}(u)$
is the seam/boundary free energy. There is no seam free energy for periodic boundary conditions without a seam. The bulk, seam and boundary
free energies can be calculated~\cite{Stroganov,Bax82Inv,OBrienP97}
by the inversion relation method. In the following, we derive 
the functional relations in the different geometries.

\subsubsection{Periodic boundary condition}

The entries of the transfer matrices $\vec T_{1,2}$ with periodic boundary conditions are defined
on a lattice with $N$ columns with $N$ even by
\begin{equation}
\vec T_{j}(u)_{\svec a}^{\svec b}=W_{j}\Big(\Big.\begin{array}{cc}
b_{1} & b_{2}\\
a_{1} & a_{2}
\end{array}\Big|u\Big)W_{j}\Big(\Big.\begin{array}{cc}
b_{2} & b_{3}\\
a_{2} & a_{3}
\end{array}\Big|u\Big)\dots W_{j}\Big(\Big.\begin{array}{cc}
b_{N-1} & b_{N}\\
a_{N-1} & a_{N}
\end{array}\Big|u\Big)W_{j}\Big(\Big.\begin{array}{cc}
b_{N} & b_{1}\\
a_{N} & a_{1}
\end{array}\Big|u\Big)
\end{equation}
For $j=1$, we take the fundamental weight $W_{1}=W$. For $j=2$, 
the transfer matrix $T_{2}(u)$ coincides with the $1\times 2$ fused face weight
\begin{equation}
W_{2}\Big(\Big.\begin{array}{cc}
d & c\\
a & b
\end{array}\Big|u\Big)=\frac{s(\lambda)}{s(u)}\sum_{e}W\Big(\Big.
\begin{array}{cc}
e & f\\
a & b
\end{array}\Big|u\Big)W\Big(\Big.\begin{array}{cc}
d & c\\
e & f
\end{array}\Big|u\!+\!\lambda\Big)
\end{equation}
In the gauge $g_a=(-1)^{a/2}$, these weights are independent of $f$ and are non-vanishing only when $|a-d|=0,2$ and $a+d=4,6$.
The transfer matrices $\vec T_{1,2}$ satisfy a simple fusion functional equation
\begin{equation}
\vec T_{1}(u)\vec T_{1}(u+\lambda)=\Big(\frac{s(u+\lambda)s(\lambda-u)}{s^{2}(\lambda)}
\Big)^{N}\mathbbm{1}+\Big(\frac{s(u)}{s(\lambda)}\Big)^{N}\vec T_{2}(u)
\end{equation}

The Lee-Yang theory is the simplest of the non-unitary RSOS models. 
In this case, the fused weights $W_{2}$ are trivially related to
$W_{1}$ modulo some $u$-independent gauge factors
\begin{equation}
W_{2}\Big(\Big.\begin{array}{cc}
d & c\\
a & b
\end{array}\Big|u\Big)=\frac{G_aG_c}{G_bG_d}\,W\Big(\Big.\begin{array}{cc}
d & c\\
5-a & 5-b
\end{array}\Big|u+3\lambda\Big)
\end{equation}
The gauge factors $G_a$ do not contribute to the transfer
matrix so that $\vec T_{1}(u)= \vec Y\vec T_{2}(u)=\vec T_{2}(u)\vec Y$. If we normalize
the transfer matrix
\begin{equation}
\vec T(u)=\Big(\frac{s(\lambda)s(u+2\lambda)}{s(u+\lambda)s(u+3\lambda)}\Big)^{N}
\vec T_{1}(u)\label{eq:pernorm}
\end{equation}
then $\vec T(u)$ satisfies the functional relation
\begin{equation}
\vec T(u)\vec T(u+\lambda)=\mathbbm{1}+\vec Y\vec T(u\!+\!3\lambda)
\end{equation}
The height reversal matrix $\vec Y$ commutes with $\vec T$ so it can diagonalized
in the same basis. Since $\vec Y^{2}=1$ the eigenvalues are $Y=\pm1$.
Restricting our analysis to the $Y=+1$ eigenspace, the eigenvalues $t(u)$
of the transfer matrix $\vec t(u)$ satisfy the functional equation
\begin{equation}
t(u)t(u+\lambda)=1+t(u+3\lambda)\label{eq:t-system}
\end{equation}
The eigenvalues of the transfer matrix also satisfies the crossing
relation and periodicity 
\begin{equation}
t(\lambda-u)=t(u),\qquad t(u+\pi)=t(u)\label{eq:crossing+periodicity}
\end{equation}

\subsubsection{Periodic boundary condition with a seam}

The entries of the transfer matrices $\vec T_{1,2}$ for periodic boundary condition and a simple single-defect seam~\cite{CMOP} with
parameter $\xi$ on a lattice of $N$ columns with $N$ even are
\begin{equation}
\vec T_{j}^{s}(u)_{\svec a}^{\svec b}=W_{j}\Big(\Big.\begin{array}{cc}
b_{1} & b_{2}\\
a_{1} & a_{2}
\end{array}\Big|u\!+\!\xi\Big)W_{j}\Big(\Big.\begin{array}{cc}
b_{2} & b_{3}\\
a_{2} & a_{3}
\end{array}\Big|u\Big)\dots W_{j}\Big(\Big.\begin{array}{cc}
b_{2N-1} & b_{2N}\\
a_{2N-1} & a_{2N}
\end{array}\Big|u\Big)W_{j}\Big(\Big.\begin{array}{cc}
b_{2N} & b_{1}\\
a_{2N} & a_{1}
\end{array}\Big|u\Big)\label{seamT}
\end{equation}
where the superscript $s$ refers to the seam. The parameter $\xi$
is arbitrary and can be complex. 
These transfer matrices satisfy the functional equation
\begin{equation}
\vec T_{1}^{s}(u)\vec T_{1}^{s}(u\!+\!\lambda)=\frac{s(u\!+\!\lambda\!+\!\xi)s(\lambda\!-\!u\!-\!\xi)}{s^{2}
(\lambda)}\Big(\frac{s(u\!+\!\lambda)s(\lambda\!-\!u)}{s^{2}(\lambda)}\Big)^{N-1}
\mathbbm{1}+\frac{s(u\!+\!\xi)}{s(\lambda)}\Big(\frac{s(u)}{s(\lambda)}\Big)^{N-1}
\vec T_{2}^{s}(u)
\end{equation}
If we restrict to the $Y=+1$ eigenspace and normalize the transfer matrix
\begin{equation}
\vec T^{s}(u)=\frac{s(\lambda)s(u+2\lambda+\xi)}{s(u+\lambda+\xi)s(u+3\lambda+\xi)}
\Big(\frac{s(\lambda)s(u+2\lambda)}{s(u+\lambda)s(u+3\lambda)}\Big)^{N-1}
T_{1}^{s}(u)
\label{seamnorm}
\end{equation}
then the eigenvalues $t(u)$ satisfy the functional relation
(\ref{eq:t-system}). 

We note that the $\xi\to0$ limit reproduces
the periodic result corresponding to the $(r,s)=(1,1)$ identity seam. To produce the $(r,s)=(1,2)$ conformal seam, we renormalize the critical seam weights and take the braid limit $\xi\to i\infty$
\begin{equation}
W^s\Big(\Big.\begin{array}{cc}
d & c\\
a & b
\end{array}\Big)=
\lim_{\xi\to i\infty} \frac{s(\lambda)}{s(\frac{\lambda}{2}\!-\!u\!-\!\xi)}\,
W\Big(\Big.\begin{array}{cc}
d & c\\
a & b
\end{array}\Big|u+\xi\Big)=e^{i\frac{\lambda}{2}}\,\delta(a,c)-
e^{-i\frac{\lambda}{2}}\,\frac{g_a}{g_c} \frac{s(c\lambda)}{s(b\lambda)}\,\delta(b,d)
\label{12seam}
\end{equation}

\subsubsection{Boundary case}

To ensure integrability of the Lee-Yang lattice model in the presence
of a boundary~\cite{BPO96} we need commuting double row transfer
matrices and triangle boundary weights which satisfy the boundary
Yang-Baxter equations. The conformal boundary conditions are labelled
by the Kac labels $(r,s)=(1,s)$ with $s=1,2$. These conformal boundary 
conditions can be realized in terms of integrable boundary conditions in different ways. 
For $(r,s)=(1,1)$, the non-zero
left and right triangle weights are given by 
\begin{eqnarray}
K_{L}\Big(\Big.\begin{array}{c}
1\\
1
\end{array}2\Big|u\Big)=\frac{s(2\lambda)}{s(\lambda)},\qquad
 K_{R}\Big(\Big.2\begin{array}{c}
1\\
1
\end{array}\Big|u\Big)=1
\end{eqnarray}
Another integrable boundary can be constructed~\cite{BP00} by
acting with a seam on the $(1,1)$ integrable boundary. The non-zero right boundary weights are
\begin{eqnarray}
K_{R}\Big(\Big.a\begin{array}{c}
2\\
2
\end{array}\Big|u,\xi\Big) & = & W\Big(\Big.\begin{array}{cc}
a & 2\\
2 & 1
\end{array}\Big|u\!+\!\xi\Big)W\Big(\Big.\begin{array}{cc}
2 & 1\\
a & 2
\end{array}\Big|\lambda\!-\!u\!+\!\xi\Big)K_{R}\Big(\Big.2\begin{array}{c}
1\\
1
\end{array}\Big|u\Big)\\[6pt]
 & = & \frac{s(u+\xi+(2\!-\!a)\lambda)s(u-\xi+(2\!-\!a)\lambda)}{s(\lambda)s(2\lambda)},
\quad a=1,3\nonumber\label{xibdy}
\end{eqnarray}
Generically, for RSOS models in the continuum scaling limit with $\xi$ real, this integrable boundary 
converges to either the $(r,s)=(1,1)$ or $(2,1)$ conformal boundary condition depending on the real value of $\xi$. 
However, since $(r,s)=(2,1)$ does not exist for the Lee-Yang theory, this boundary condition converges in the continuum scaling limit to the $(1,1)$ conformal boundary condition for all real values of $\xi$.
To obtain the $(r,s)=(1,2)$ boundary condition, we need to renormalize these boundary weights to obtain a nontrivial braid limit as $\xi\to i\infty$. The explicit $(r,s)=(1,2)$ non-zero right boundary weights in this limit are 
\begin{eqnarray}
 K_{R}\Big(\Big.1\begin{array}{c}
2\\ 2
\end{array}\Big|u\Big)=K_{R}\Big(\Big.3\begin{array}{c}
2\\ 2
\end{array}\Big|u\Big)=1
\end{eqnarray}
In this way, varying $\Im\xi$ between $0$ and $\infty$ in (\ref{xibdy}) interpolates between the $(1,1)$ and $(1,2)$ conformal boundary conditions. 
Integrability, in the presence of these boundaries, derives from the fact that these
boundary weights satisfy the left and right boundary Yang-Baxter equations,
respectively. In the following, we fix the left boundary weight to be 
$(1,1)$ and take the right boundary weight to be either $(1,1)$ or
$(1,2)$. 

The face and triangle boundary weights are used to construct~\cite{BPO96}
a family of commuting double row transfer matrices $\vec T^{b}(u)$, where
$b$ refers to the boundary case. For a lattice of width $N$, the entries
of the transfer matrix $\vec T^{b}(u)$ are
\begin{eqnarray}
\vec T_{1}^{b}(u)_{\svec a}^{\svec b} & = & \sum_{c_{1},\dots,c_{N}}
K_{L}\Big(\Big.\begin{array}{c}
1\\
1
\end{array}2\Big|\lambda\!-\!u\Big)W\Big(\Big.\begin{array}{cc}
1 & b_{1}\\
2 & c_{1}
\end{array}\Big|\lambda\!-\!u\Big)W\Big(\Big.\begin{array}{cc}
b_{1} & b_{2}\\
c_{1} & c_{2}
\end{array}\Big|\lambda\!-\!u\Big)\dots W\Big(\Big.\begin{array}{cc}
b_{N-1} & s\\
c_{N-1} & c_N
\end{array}\Big|\lambda\!-\!u\Big)\nonumber\\
& & \mbox{}\times 
 W\Big(\Big.\begin{array}{cc}
2 & c_{1}\\
1 & a_{1}
\end{array}\Big|u\Big)W\Big(\Big.\begin{array}{cc}
c_{1} & c_{2}\\
a_{1} & a_{2}
\end{array}\Big|u\Big)\dots W\Big(\Big.\begin{array}{cc}
c_{N-1} & c_N\\
a_{N-1} & s
\end{array}\Big|u\Big)K_{R}\Big(\Big.c_{N}\begin{array}{c}
s\\
s
\end{array}\Big|u\Big)
 \quad\qquad\label{RTMdef}
\end{eqnarray}
 This transfer matrix is positive definite and satisfies crossing
symmetry $\vec T_{1}^{b}(u)=\vec T_{1}^{b}(\lambda-u)$. Although $\vec T_{1}^{a}(u)$
is not symmetric or normal, this one-parameter family of transfer
matrices can be diagonalized because $\tilde{\vec T}_{1}^{b}(u)=\vec G\vec T_{1}^{b}(u)=
\tilde{\vec T}_{1}^{b}(u)^{T}$
is symmetric where the diagonal gauge matrix $\vec G$ is given by 
\begin{eqnarray}
\vec G_{\svec a}^{\svec b}=\prod_{j=1}^{N-1}G(a_{j},a_{j+1})\,\delta(a_{j},b_{j}) 
& \quad\mbox{with}\quad & G(a,b)=\begin{cases}
s(\lambda)/s(2\lambda), & b=1,4\\
1, & \mbox{otherwise}
\end{cases}
\end{eqnarray}
 It is convenient to define the normalized transfer matrix 
\begin{equation}
\vec T^{b}(u)=S_{b}(u)\,\frac{s^{2}(2u-\lambda)}{s(2u+\lambda)s(2u-3\lambda)}
\Bigl(\frac{s(\lambda)s(u+2\lambda)}{s(u+\lambda)s(u+3\lambda)}\Bigr)^{2N}\vec T(u)
\label{eq:bdrynorm}
\end{equation}
 with 
\begin{eqnarray}
S_{s}(u) & = & \begin{cases}
1, & s=1\\
\frac{s^{2}(\lambda)}{s(u+\xi+\lambda)s(u-\xi+\lambda)s(u+\xi-\lambda)
s(u-\xi-\lambda)} & s=2
\end{cases}
\end{eqnarray}
It can then be shown~\cite{BPO96} that, in the $Y=+1$ eigenspace, the eigenvalue $t(u)$ of the normalized
transfer matrix $\vec T(u)=\vec T_1(u)$ satisfies the universal $Y$-system
(\ref{eq:t-system}) independent of the boundary conditions labelled by $s=1,2$.

\section{Classification of States}

In this section, we exhaustively classify the states of the Lee-Yang theory. We start with
the critical case and describe correspondences between the conformal
basis~\cite{FNO}, the distribution of complex zeros of the transfer matrix eigenvalue
and certain RSOS paths~\cite{FodaW,FevPW} related to the one-dimensional configurational sums which appear in Baxter's Corner Transfer Matrices (CTMs)~\cite{BaxterCTM}.
After analysing the classification for finite $N$ in the
various geometries, we comment on the behaviour of finite excitations above the ground state 
in the continuum scaling limit and their off-critical counterparts.

\subsection{$(m,n)$ systems, zero patterns, RSOS paths and characters}

Consider the critical Lee-Yang lattice model (\ref{RSOS}) with $\lambda=\frac{3\pi}{5}$ and 
$s(u)=\sin(u)$. To describe the correspondence with the conformal Lee-Yang
model with central charge $c=-\frac{22}{5}$, we first recall the description of the Lee-Yang Virasoro modules. For this theory, the Virasoro algebra admits two irreducible modules with characters
\begin{equation}
\chi_h(q)=q^{-\frac{c}{24}+h}\sum_{n=0}^{\infty}\mbox{dim}(V_{n}^{h})\,q^{n},\qquad h=0,-\tfrac{1}{5}
\end{equation}
where $n=E$ is the energy ($L_{0}$ eigenvalue) of the given state. 
The identity $h=0$ module, is built~\cite{FNO,Fortin:2004ct} over the vacuum $\vert0\rangle$ by the states 
\begin{equation}
L_{-n_{1}}\dots L_{-n_{m}}\vert0\rangle,\qquad n_{m}>1,\qquad
 n_{i}>n_{i+1}+1\label{eq:V0basis}
\end{equation}
Due to the constraint on the indices, this basis is linearly independent and 
there are no singular vectors. This representation space has the reduced 
Virasoro character \cite{Nahm:1992sx}
\begin{equation}
\hat\chi_{0}(q)=q^{\frac{c}{24}}\chi_0(q)=
\sum_{n=0}^{\infty}\frac{q^{n^{2}+n}}{(1-q)\dots(1-q^{n})}=
\prod_{n=1}^{\infty}\frac{1}{(1-q^{5n-3})(1-q^{5n-2})}\label{eq:chi0}
\end{equation}
The sum and product forms are related by the Andrews-Gordon identity 
which is a generalization of the Rogers-Ramanujan identities. 
The second module is built over the highest weight state
$\vert h\rangle$ with conformal weight $h=-\frac{1}{5}$. This module is generated
by the linearly independent modes 
\begin{equation}
L_{-n_{1}}\dots L_{-n_{m}}\vert h\rangle,\qquad n_{m}>0,\qquad
 n_{i}>n_{i+1}+1\label{eq:V1basis}
\end{equation}
and has the reduced Virasoro character
\begin{equation}
\hat\chi_{-\frac{1}{5}}(q)=q^{\frac{c}{24}+\frac{1}{5}}\chi_{-\frac{1}{5}}(q)=
\sum_{n=0}^{\infty}\frac{q^{n^{2}}}{(1-q)\dots(1-q^{n})}=
\prod_{n=1}^{\infty}\frac{1}{(1-q^{5n-4})(1-q^{5n-1})}
\label{eq:chi1}
\end{equation}

The Hilbert space of the Lee-Yang lattice model can be viewed as a space of RSOS paths. The entries of the unrenormalized transfer matrix are Laurent polynomials in the $z=e^{iu}$
and $z^{-1}=e^{-iu}$ of finite degree determined by $N$. Because the transfer matrices are commuting families, with a common set of $u$-independent eigenvectors, it follows that the eigenvalues are also Laurent polynomials of the same degree. These polynomials can be obtained by explicit numerical diagonalization and their zeros obtained by numerical factorization. The various eigenvalues are thus characterized
by the location and pattern of the zeros in the complex $u$-plane. We will describe the relation 
between the RSOS paths, the patterns of zeros and the two Virasoro
bases given by (\ref{eq:V0basis}) and (\ref{eq:V1basis}) in each of the
space-time geometries. We start with the simplest case which is the boundary case. 
In this case, the Hilbert space consists of a single Virasoro module. 
We next analyse the periodic case with and without the
seam, where tensor products of left and right chiral Virasoro modules appear. Lastly, we analyse
the flows between the modules induced by increasing $\Im \xi$ from $0$ to $\infty$.

\subsubsection{Boundary case}

\begin{figure}
\begin{centering}
\includegraphics[height=6cm]{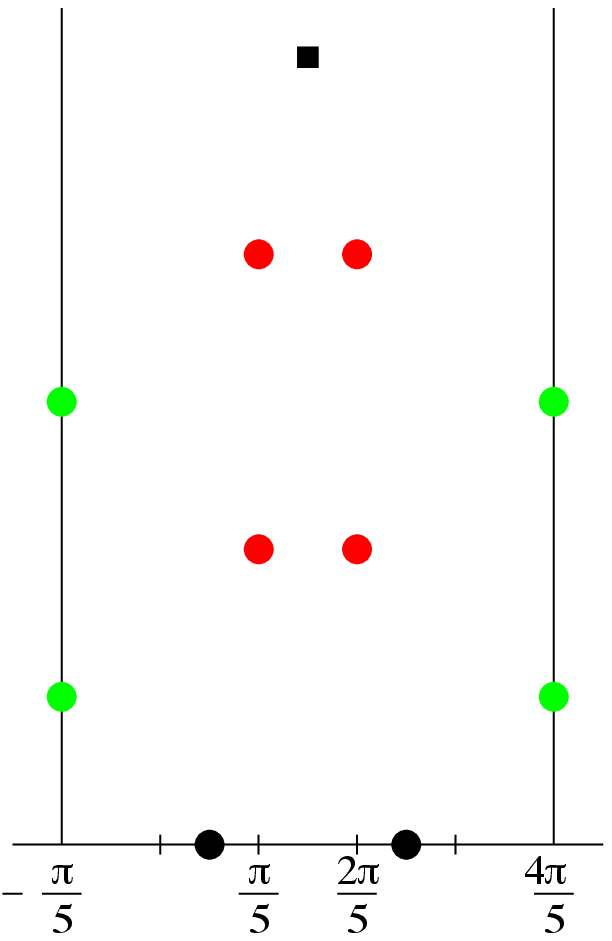}~~~~~~~\includegraphics[height=6cm]{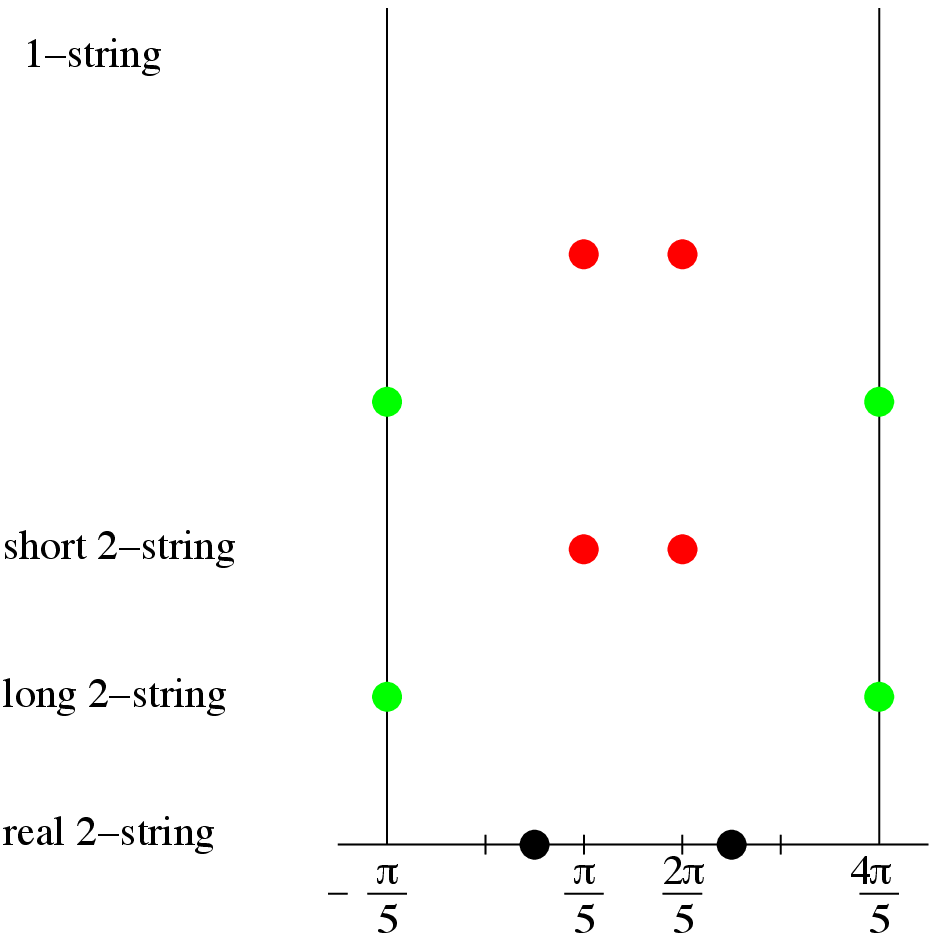}
\par\end{centering}

\caption{\label{typicalConfig}A typical pattern of zeros of a transfer matrix eigenvalue. The four kinds of strings are shown in the analyticity strip within the upper-half complex $u$-plane. The $(1,1)$ boundary condition is on the left and $(1,2)$ on the right. A single $1$-string (shown in black) occurs furthest from the real axis with a fixed location for all eigenvalues in the $(1,1)$ sector. No short $2$-strings (shown in red) occur in the ground states for which the Fermi sea of long $2$-strings (shown in green) is full. A real 2-string (shown in black) occurs on the real axis.}
\end{figure}

\paragraph{$(m,n)$ systems and zero patterns.}
Let us consider the sectors with $(r,s)=(1,1),(1,2)$ boundary conditions
which we label simply by $s=1,2$. The excitation energies are
given by the eigenvalues of the double-row transfer matrix $\vec T^{b}(u)$ where
$u$ is the spectral parameter. The single relevant analyticity strip
in the complex $u$-plane is the full periodicity strip 
\begin{equation}
-\frac{\pi}{5}<\Re u<\frac{4\pi}{5}
\end{equation}
In this case, the transfer matrix is symmetric under complex conjugation so it suffices to analyse the eigenvalue zeros on the upper half plane. The
zeros form strings and the excitations are classified by the string
content in the analyticity strip. Four different kinds of strings occur 
which we designate ``$1$-strings'', ``short $2$-strings'',
``long $2$-strings'' and ``real $2$-strings'' as indicated in Figure~\ref{typicalConfig}. 

A $1$-string $u_{j}=3\pi/10+iv_{j}$ lies in the middle of the
analyticity strip and has real part $3\pi/10$. It occurs only in the $(1,1)$
sector and has a fixed location for all eigenvalues in this sector. 
A short $2$-string consists of a pair of zeros
$u_{j}=\pi/5+iy_{j}$, $2\pi/5+iy_{j}$ with common imaginary parts and real parts $\pi/5$,
$2\pi/5$ respectively. The two zeros $u_{j}=-\pi/5+iy_{j}$,
$4\pi/5+iy_{j}$  of a long $2$-string have common imaginary parts and real parts $-\pi/5$,
$4\pi/5$ respectively so that these zeros sit at the edge of the
analyticity strip. Strictly speaking, due to periodicity, long $2$-strings could be considered as $1$-strings. 
Here, however, we follow the nomenclature used for RSOS models with more than one analyticity strip. 
Lastly, a real 2-string consists of a pair of zeros $u_{j}=\pi/10,\pi/2$
on the real axis. Usually the real parts of such strings approach
these special values for large $N$ but here, because of symmetries,
the values of these real parts are exact for finite $N$. 

The string contents are described by $(m,n)$ systems~\cite{Melzer,Berk94} which, for the Lee-Yang model, take the form
\begin{equation}
\begin{array}{rl}
 &\ \  2m+n+3-s=N\ \ \Leftrightarrow\ \  m+n=N-m+s-3,\qquad s=1,2 \\
 & m=\text{\{number of short \ensuremath{2}-strings\}},\qquad n=
\text{\{number of long \ensuremath{2}-strings\}}
\end{array}\label{stringdefs}
\end{equation}
There is always a real 2-string on the real axis and, in the $(r,s)=(1,1)$
sector, a single $1$-string furthest from the real axis. Each short
$2$-string contributes two zeros and, by periodicity,
each long $2$-string contributes one zero. The 1-string contributes
one zero and so does the real 2-string since it is shared between
the upper and lower half planes. Consequently, the $(m,n)$ system
expresses the conservation of the $2N$ zeros in the periodicity
strip. No short $2$-strings occur in the ground states for which the ``Fermi sea'' of long $2$-strings is full. 
For finite excitations above the ground state, $m$ is finite but $n\sim N$ as
$N\to\infty$.

An excitation with string content $(m,n)$ is uniquely labelled by
a set of quantum numbers
\begin{equation}
I=(I_{1},I_{2},\ldots,I_{m})
\end{equation}
 where the integers $I_{j}\ge0$ give the number of long 2-strings
whose imaginary parts are greater than that of the given
short 2-string $y_{j}$. The short 2-strings and long 2-strings labelled by $j=1$ 
are closest to the real axis. The quantum
numbers $I_{j}$ satisfy 
\begin{equation}
n\ge I_{1}\ge I_{2}\ge\dots\ge I_{m}\ge0.\label{iranges}
\end{equation}
 For given string content $(m,n)$, the lowest excitation occurs when
all of the short 2-strings are further out from the real axis than
all of the long 2-strings. In this case all of the quantum numbers
vanish $I_{j}=0$. Bringing the location of a short 2-string closer
to the real axis by interchanging the location of the short 2-string
with a long 2-string increments its quantum number by one unit and
increases the energy.

\paragraph{Finitized characters.}
On the lattice it is convenient to work with finitized versions of the Virasoro characters
(\ref{eq:chi0}) and (\ref{eq:chi1}) which converge to them in the $N\to\infty$
limit. These finitized characters are fermionic in the sense that they are given by $q$-polynomials with nonnegative integer coefficients.
The finitized characters are 
\begin{eqnarray}
(r,s)=(1,1):\ \ &  & \hat\chi_{0}^{(N)}(q)=\sum_{m=0}^{\floor{N/2}}q^{m^{2}+m}\Big[\begin{array}{c}
N\!-\!1\!-\!m\\
m
\end{array}\Big]_{q}\to\hat\chi_{0}(q)\\
(r,s)=(1,2): \ \ &  & \hat\chi_{-1/5}^{(N)}(q)=\sum_{m=0}^{\floor{N/2}}q^{m^{2}}\Big[\begin{array}{c}
N\!-\!m\\
m
\end{array}\Big]_{q}\,\to\,\hat\chi_{-1/5}(q)
\end{eqnarray}
where the $q$-binomial satisfies
\begin{eqnarray}
\Big[\begin{array}{c}
N\\
m
\end{array}\Big]_{q}=\,\prod_{i=1}^{m}\frac{1-q^{N+1-i}}{1-q^{i}}
\to\,\prod_{i=1}^{m}\frac{1}{1-q^{i}},\qquad N\to\infty
\end{eqnarray}
Setting $q=1$ gives the correct counting of states in terms of binomial coefficients
\begin{eqnarray}
\hat\chi_{0}^{(N)}(1)=\!\!\sum_{m=0}^{\floor{N/2}}\Big(\begin{array}{c}
N\!-\!1\!-\!m\\
m
\end{array}\Big)=F_{N},\quad\ \ 
\hat\chi_{-1/5}^{(N)}(1)=\!\sum_{m=0}^{\floor{N/2}}\Big(\begin{array}{c}
N\!-\!m\\
m
\end{array}\Big)=F_{N+1}\quad
\end{eqnarray}
where the Fibonacci numbers are given by
\begin{equation}
F_N=F_{N-1}+F_{N-2}=1,1,2,3,5,8.\ldots,\qquad N=1,2,3,4,5,6,\ldots
\end{equation}

The finitized characters can also be written in the form of 1-dimensional configurational sums related to Baxter's CTMs
\begin{eqnarray}
\hat\chi_{h}^{(N)}(q)=\sum_E q^{E}=\sum_{\sigma}q^{\sum_{j=1}^{N}jH(\sigma_{j-1},
\sigma_{j},\sigma_{j+1})}\label{eq:fincharbdry}
\end{eqnarray}
where the first sum is over integer conformal energies $E$ and the second sum is over all one-dimensional RSOS paths $\sigma=\{\sigma_{0},
\sigma_{1},\ldots,\sigma_{N}\}$
on $A_{4}$ with $\sigma_{0}=s$ and either $(\sigma_{N},\sigma_{N+1})=(2,3)$ or $(\sigma_{N},\sigma_{N+1})=(3,2)$ depending on the parity of $N$. The \emph{energy
function} $H$ is 
\begin{equation}
H(\sigma_{j-1},\sigma_{j},\sigma_{j+1})=\begin{cases}
1, & (\sigma_{j-1},\sigma_{j},\sigma_{j+1})=(2,1,2)\ \mbox{or}\ (3,4,3)\\
0, & \text{otherwise}
\end{cases}
\end{equation}
This local energy function differs from, but is gauge equivalent to, the one used in~\cite{FB85}. 
The lowest configuration energy $E=0$ occurs for a ground state RSOS path with heights alternating between 2 and 3.

\paragraph{Bijection between zero patterns and RSOS paths.}
\begin{figure}[t]
\label{bijection}
\begin{centering}
\raisebox{.3cm}{\includegraphics[height=5cm]{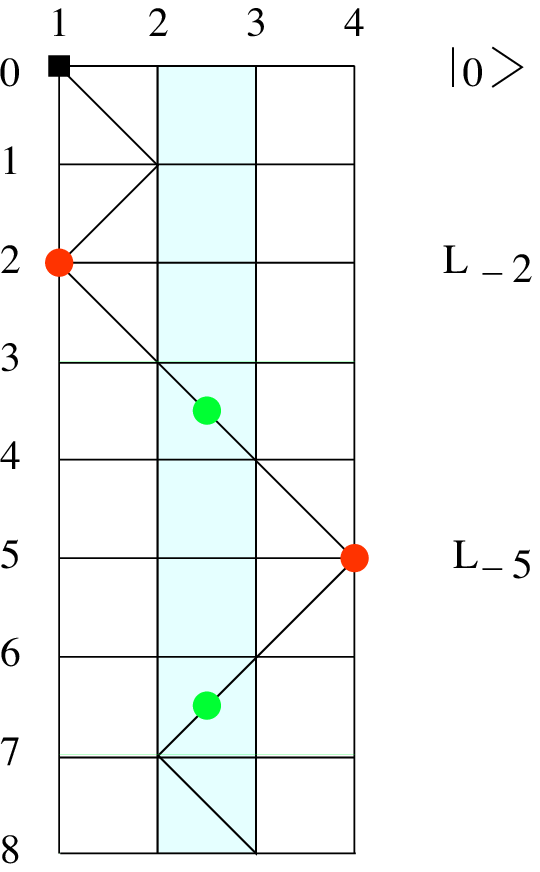}}~~~\includegraphics[height=5cm]{boundzero1New}
~~~~~~~\raisebox{.3cm}{\includegraphics[height=5cm]{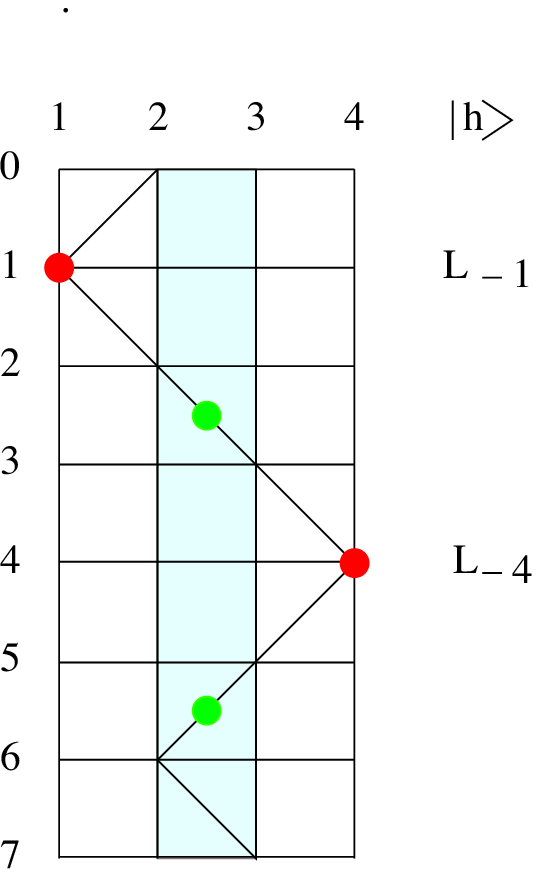}}~~~\includegraphics[height=5cm]{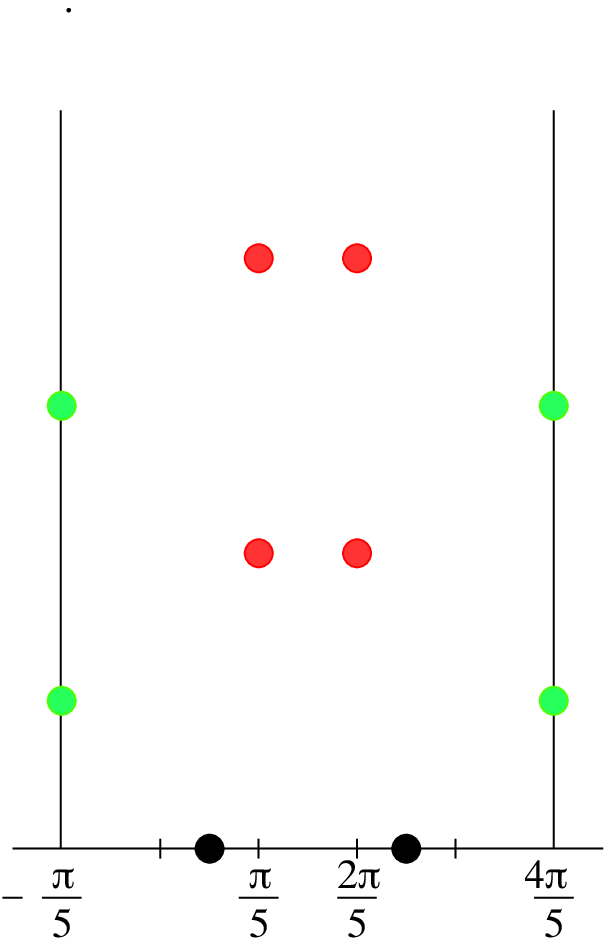}
\par\end{centering}
\vspace{-0.1in}
 \caption{\label{bijection}On the left is shown the bijection between one-dimensional configurational paths $\sigma$, strings in the periodicity strip and the Virasoro
descendants over $\vert0\rangle$ or $|h\rangle=|\Phi\rangle=|\!-\frac{1}{5}\rangle$. The paths $\sigma$ are rotated $90^{\circ}$ clockwise and finish with height 2 followed by height 3. The ground states are given by alternation between heights 2 and 3 in the shaded band. 
On the right is shown the associated patterns of zeros in the analyticity strip. 
An $(r,s)=(1,1)$ example is shown 
on the left and an $(r,s)=(1,2)$ example on the right.
The $1$-string (solid black square),
short $2$-strings (solid red disks) and long $2$-strings (solid green disks) in
the upper-half complex $u$-plane are shown maintaining their relative vertical positions. 
A short 2-string at position $j$ is associated with the Virasoro mode $L_{-j}$ at the given vertical position. 
For the $(1,1)$ case, the string (particle) content is $m=n=2$ with
$N=8$ and modes inserted at $j=2,5$. For the $(1,2)$ case, the string (particle) content is $m=n=2$ with $N=7$ and modes inserted at $j=1,4$.
The energies of these excited states are $E=2+5=7$ and $E=1+4=5$ respectively. The corresponding  Virasoro sates are $L_{-5}L_{-2}\vert0\rangle$ and $L_{-4}L_{-1}\vert h\rangle$ respectively.}
\end{figure}
There is in fact a bijection~\cite{FP02,FPW} between the allowed patterns of strings 
in the periodicity strip, the one-dimensional 
RSOS paths $\sigma$ that label the eigenstates (eigenvalues)
and Virasoro descendant states. Although this bijection is useful for counting 
and classifying states, it does not imply that each individual eigenstate is to be identified 
with the individual Virasoro state. 
Indeed, there are typically many degenerate levels in any energy eigenspace 
and, in general, a given eigenstate can only be written as a linear combination with other linearly independent Virasoro states at that energy level.
 However, the independent eigenstates and Virasoro states are equinumerous within each fixed energy eigenspace.

A simple natural bijection is constructed as follows. A consecutive pair of heights 
$(\sigma_{j},\sigma_{j+1})=(2,3)$ or $(3,2)$ corresponds
to a long $2$-string (dual particle) at position $j+\frac{1}{2}$. 
The alternation of heights between $2$ and $3$ give the lowest energy 
configuration and the ground state $|h\rangle$ in a given sector.
A triple $(\sigma_{j-1},\sigma_{j},\sigma_{j+1})=(2,1,2)$
or $(3,4,3)$ corresponds to a short $2$-string (particle) at position
$j$ and an insertion of a Virasoro mode $L_{-j}$. Because of the RSOS 
constraints, the particles obey nearest-neighbour exclusion. 
In addition, in the sector $(r,s)=(1,1)$, there is a single $1$-string
at $j=0$ (furthest from the real axis) corresponding to the initial height $s=1$ at $j=0$. 
This bijection is illustrated in Figure~\ref{bijection}. Notice that
only the relative positions of the long and short $2$-strings are 
important. In both cases, the first height is $\sigma_0=s$ 
and the last pair is $(\sigma_N,\sigma_{N+1})=(2,3)$ with the parity of $N$
fixed accordingly. A particle has an effective diameter of two units whereas a dual particle
has a diameter of one unit. 
We see that the geometric packing constraint 
\begin{equation}
2m+n+3-s=N
\end{equation}
agrees with the $(m,n)$ system.

\paragraph{Flow between boundary conditions.}
Using the boundary conditions (\ref{xibdy}), a Renormalization Group (RG) boundary flow is induced from the identity conformal boundary fixed point $\mathbbm{1}=(1,1)$
to the $\Phi=(1,2)$ conformal boundary fixed point as $\Im \xi$ increases from $0$ to $\infty$. 
The boundary entropies~\cite{AffleckLudwig} are given by~\cite{DPTW,DRTW}
\begin{eqnarray}
g_{\mathbbm{1}}=\Big(\frac{\sqrt{5}-1}{2\sqrt{5}}\Big)^{\!\frac{1}{4}}\!\!,\ \  
g_{\Phi}=\Big(\frac{2+\sqrt{5}}{\sqrt{5}}\Big)^{\!\frac{1}{4}}\!\!,\ \  
\log g_{\mathbbm{1}}^2=-0.642965,\ \ 
\log g_{\mathbbm{1}}g_{\Phi}=-0.161754\quad
\end{eqnarray}
The physical flow is from the $\Phi$ boundary fixed point to the identity $\mathbbm{1}$ fixed point so that the boundary entropy decreases along the flow. The mathematical flow we describe is in the opposite direction. 
It can be described simply in terms of each of the three different descriptions of the states, namely, the zero patterns, the RSOS paths or the Virasoro states.

\begin{figure}[t]
\begin{centering}
\includegraphics[height=4.8cm]{boundzero1New} ~~~~~~~~\includegraphics[height=6cm]{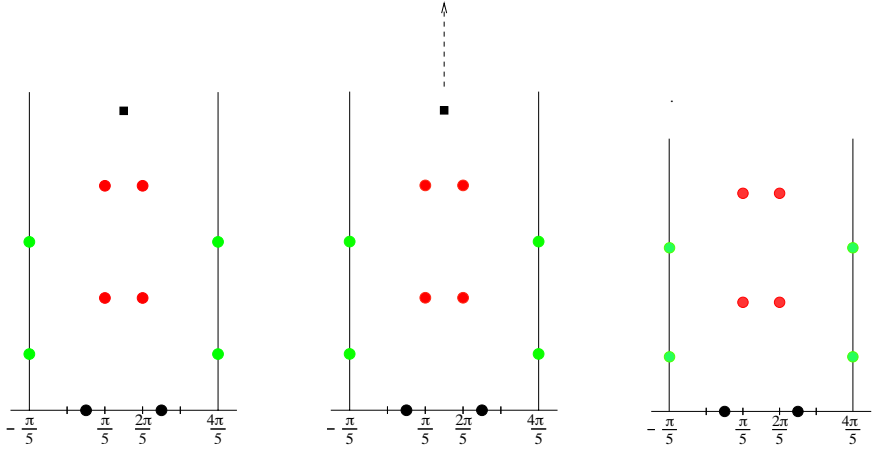}
~~~~~~~~\includegraphics[height=4.8cm]{boundzero2aNew}
\par\end{centering}

\caption{\label{bdyFlowFig}The boundary flow from $\mathbbm{1}=(1,1)$
to $\Phi=(1,2)$ in terms of the flow of zeros. On the left $\xi=0$ while
on the right $\xi=i\infty$. For intermediate values of $\xi$, shown in
the middle, the 1-string moves continuously to $i\infty$ and disappears. The relative ordering of the
2-strings is not effected. }
\end{figure}

In terms of the zeros and $(m,n)$ systems, the flow is very simple:
the 1-string which only exists for the $(1,1)$ integrable boundary condition
migrates vertically off to $i\infty$ as indicated in Figure~\ref{bdyFlowFig}. There is no change in the relative ordering of the 2-string content. 
The flow in terms of the paths is also simple. We merely have to remove
the first step of each of the $(1,1)$ paths to obtain a $(1,2)$ path. 
More interesting is the flow in terms of the Virasoro modes. First
of all, there is a flow $\vert0\rangle\mapsto\vert\Phi\rangle$ between the highest weight states. 
For a Virasoro state, the simple rule is to increase the
index of each Virasoro mode by one $L_{-n}\mapsto L_{-n+1}$:
\begin{equation}
L_{-n_{1}}L_{-n_{2}}...L_{-n_{k}}\vert0\rangle\rightarrow L_{-n_{1}+1}
L_{-n_{2}+1}...L_{-n_{k}+1}\vert \Phi\rangle
\end{equation}
This flow is summarized for the first few excited states
in Table~\ref{bdyFlowTab}. As expected and shown in the table, the flow interpolates between the 
reduced characters
\begin{align}
\hat\chi_{0}(q)&=1+q^{2}+q^{3}+q^{4}+q^{5}+2q^{6}+2q^{7}+3q^{8}+3q^{9}+...\\[4pt]
\hat\chi_{-\frac{1}{5}}(q)&=1+q+q^{2}+q^{3}+2q^{4}+2q^{5}+3q^{6}+3q^{7}+...
\end{align}
The level by level flow agrees with the TCSA results of Dorey et al.~\cite{DPTW}.

\begin{center}
\begin{table}[ht]
\begin{centering}
\begin{tabular}{|c|c|c|c|}
\hline 
Level & State in the $(1,1)$ module & State in the $(1,2)$ module & Level\tabularnewline
\hline 
\hline 
h.w. state & $|0\rangle$ & |$\Phi\rangle$ & h.w. state\tabularnewline
\hline 
2 & $L_{-2}|0\rangle$ & $L_{-1}|\Phi\rangle$ & 1\tabularnewline
\hline 
3 & $L_{-3}|0\rangle$ & $L_{-2}|\Phi\rangle$ & 2\tabularnewline
\hline 
4 & $L_{-4}|0\rangle$ & $L_{-3}|\Phi\rangle$ & 3\tabularnewline
\hline 
5 & $L_{-5}|0\rangle$ & $L_{-4}|\Phi\rangle$ & 4\tabularnewline
\cline{1-3}
6 & $L_{-2}L_{-4}|0\rangle$ & $L_{-1}L_{-3}|\Phi\rangle$ & 4\tabularnewline
\cline{3-4} 
6 & $L_{-6}|0\rangle$ & $L_{-5}|\Phi\rangle$ & 5\tabularnewline
\cline{1-2} 
7 & $L_{-2}L_{-5}|0\rangle$ & $L_{-1}L_{-4}|\Phi\rangle$ & 5\tabularnewline
\cline{3-4} 
7 & $L_{-7}|0\rangle$ & $L_{-6}|\Phi\rangle$ & 6\tabularnewline
\cline{1-2} 
8 & $L_{-2}L_{-6}|0\rangle$ & $L_{-1}L_{-5}|\Phi\rangle$ & 6\tabularnewline
8 & $L_{-3}L_{-5}|0\rangle$ & $L_{-2}L_{-4}|\Phi\rangle$ & 6\tabularnewline
\cline{3-4} 
8 & $L_{-8}|0\rangle$ & $L_{-7}|\Phi\rangle$ & 7\tabularnewline
\cline{1-2} 
9 & $L_{-2}L_{-7}|0\rangle$ & $L_{-1}L_{-6}|\Phi\rangle$ & 7\tabularnewline
9 & $L_{-3}L_{-6}|0\rangle$ & $L_{-2}L_{-5}|\Phi\rangle$ & 7\tabularnewline
\cline{3-4} 
9 & $L_{-9}|0\rangle$ & $L_{-8}|\Phi\rangle$ & 8\tabularnewline
\hline 
\end{tabular}
\par\end{centering}

\caption{\label{bdyFlowTab}The boundary flow induced by $\xi\to i\infty$ described state
by state in increasing energy.}
\end{table}
\end{center}

\subsubsection{Periodic case}

\begin{figure}[t]
\begin{centering}
\includegraphics[height=8cm]{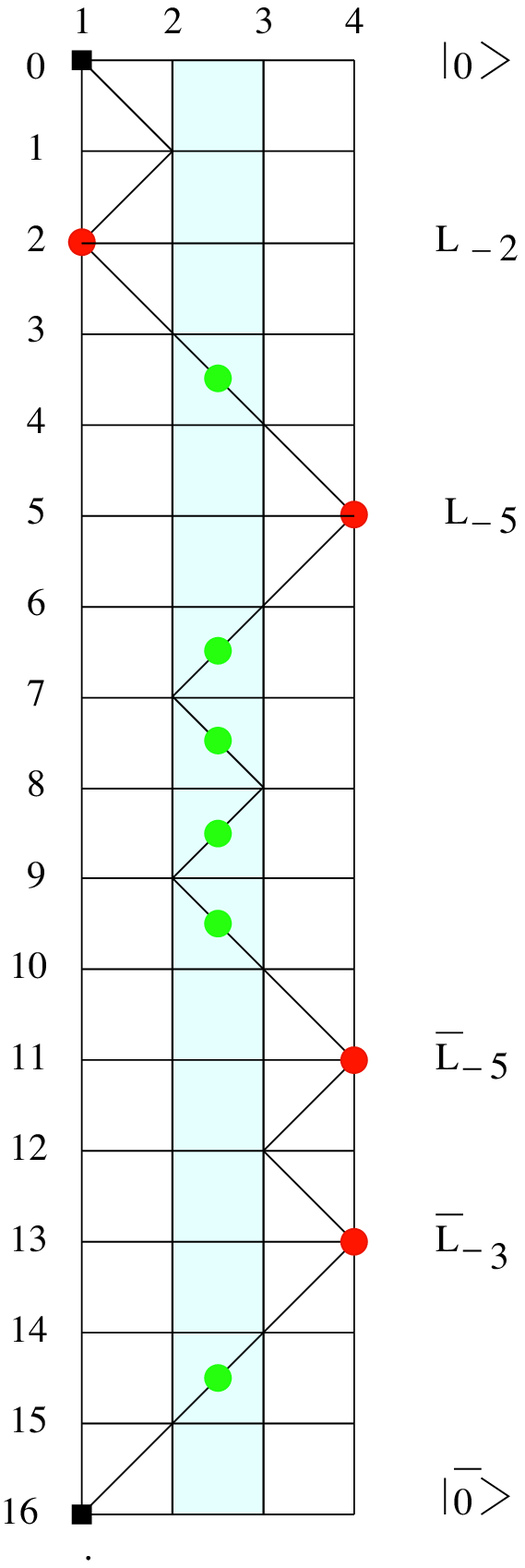}~~~\includegraphics[height=8cm]{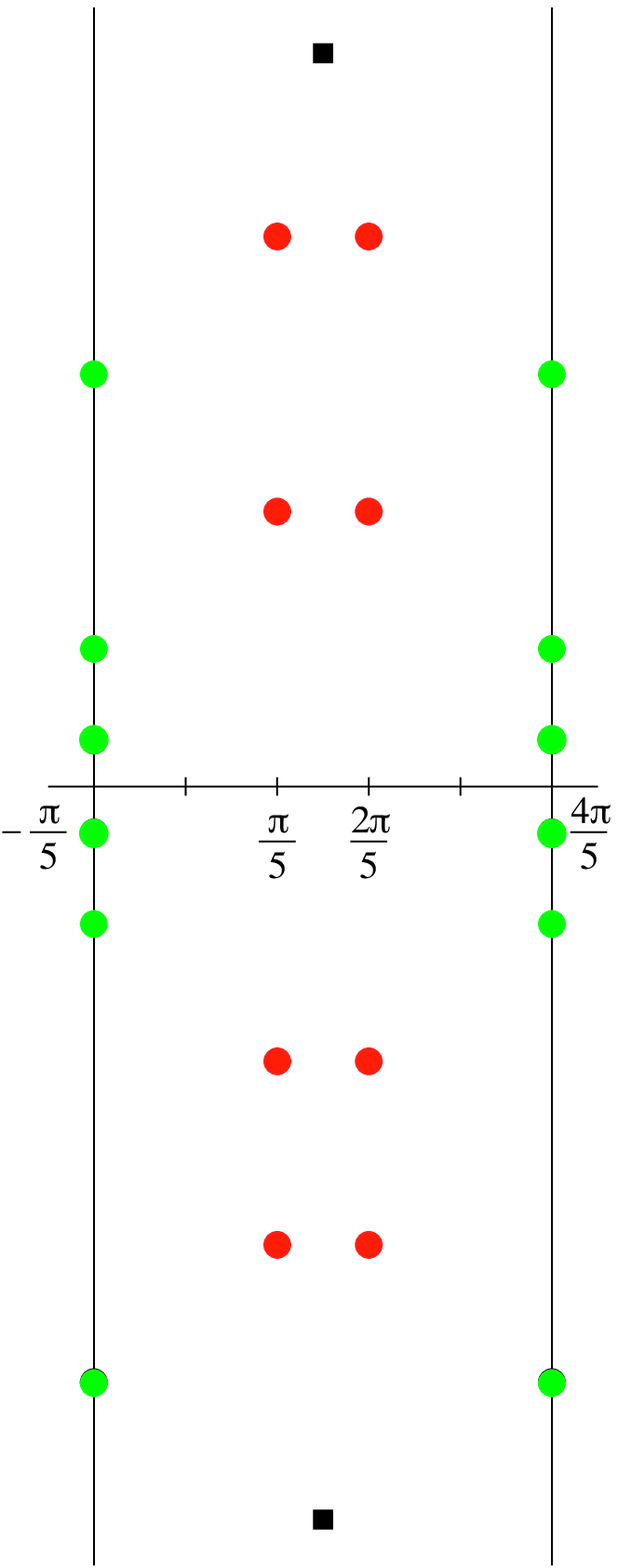}
~~~~~~~~~~~~\includegraphics[height=8cm]{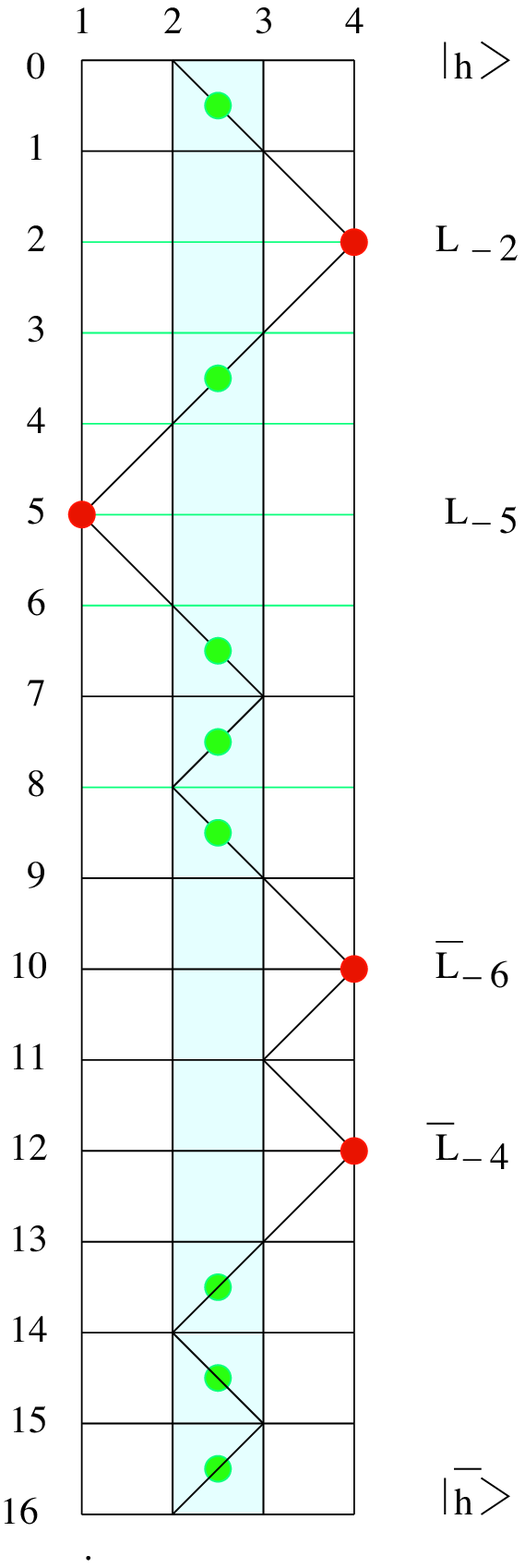}~~~\includegraphics[height=8cm]{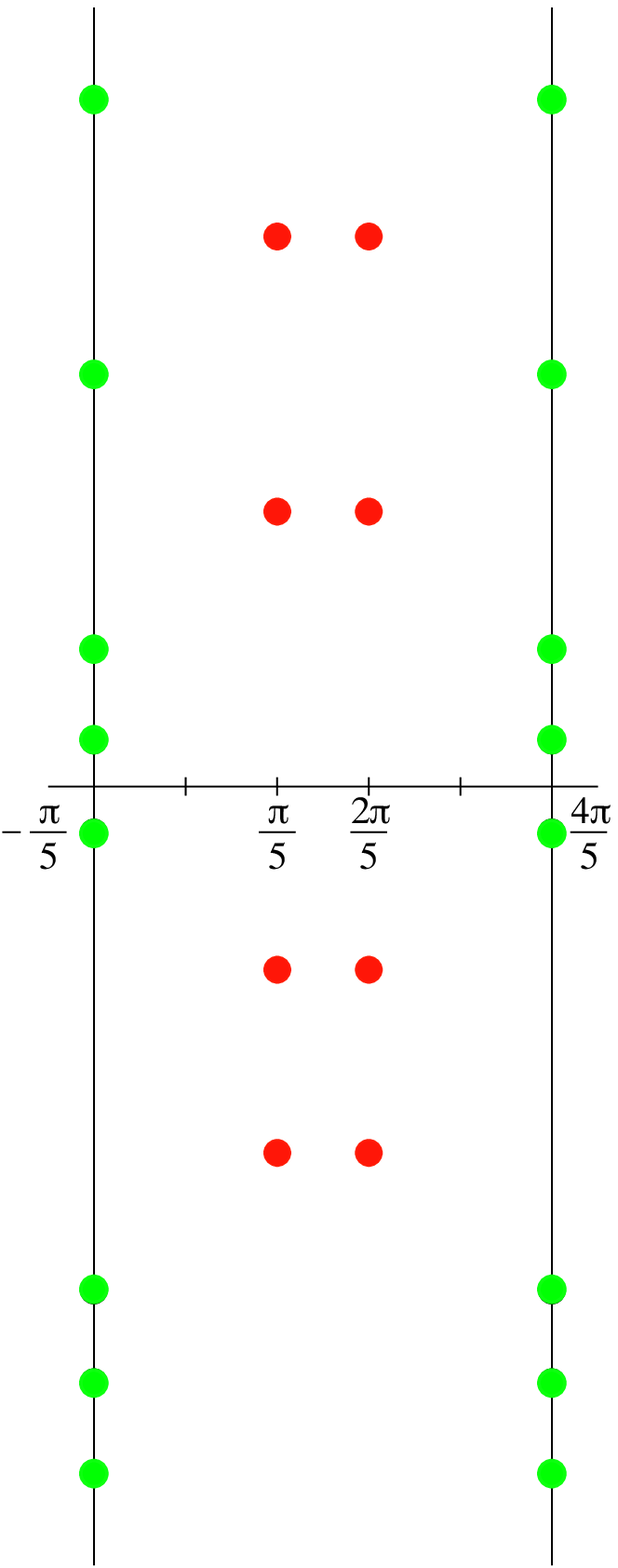}
\par\end{centering}

\vspace{-0.1in}
\caption{Two typical patterns of zeros for transfer matrix eigenvalues for $N=8$ 
in the periodic case in the sectors $(0,0)$ and $(-\frac{1}{5},-\frac{1}{5})$ respectively showing the 
identification between the zeros of the transfer matrix, paths
and Virasoro descendants. The ground state alternating band of heights 2 and 3 at the centre can 
be made arbitrarily long without effecting the energies of the eigenvalues. 
The paths start at height $s$ in the upper half-plane and finish at height $\bar{s}$ in the lower half-plane. 
In accord with the diagonal modular invariant, there is a sector selection rule that imposes $s=\bar{s}=1,2$. The cases $s=\bar{s}=3,4$ are equivalent by the $\mathbb{Z}_2$ symmetry of the $A_4$ diagram.}
\label{periodiccorrespondence} 
\label{periodicPathsZeros}
\end{figure}

For periodic boundary conditions, the excitations are again classified by the patterns of zeros. 
The known diagonal modular invariant partition function is
\begin{equation}
Z(q)=|\chi_0(q)|^2+|\chi_{-\frac{1}{5}}(q)|^2
\end{equation}
Consequently, the states separate into two sectors 
$(h,\bar{h})=(0,0)$ with $s=\bar{s}=1$ and $(h,\bar{h})=(-{1\over 5},-{1\over 5})$ with $s=\bar{s}=2$ according to 
whether 1-strings occur at the top and bottom of the analyticity strip or not as shown in Figure~\ref{periodicPathsZeros}. 
If there is a 1-string in the lower half-plane then there is always a 1-string in the upper half-plane. 
The same 1-strings and short and long 2-strings occur as
in the boundary case. In the period case, however, there are no real 2-strings on the real axis 
although a short 2-string can occur on the real axis. 
In accord with the diagonal modular invariant, there is a sector selection rule that imposes $s=\bar{s}=1,2$.
The main difference in the periodic case, is
that the zeros in the lower half plane are not necessarily related
to those in the upper half plane except for the fixed 1-strings. 
It is observed that the patterns of zeros in the upper and lower half-planes occur independently of each other and that each is described by an $(m,n)$ system.
The patterns of zeros in the upper and lower half-planes relate to the two chiral halves of the theory.

To classify the states for finite $N$, we use the classification already developed 
for the boundary case, taking into account
that the lower- and upper-half planes are independent, except for a possible
real 2-string on the real axis. By convention, this is regarded as occurring in the lower-half plane.
For the $(m,n)$ structure, we differentiate between the structures
in the upper- and lower-half planes and introduce
a doubled $(m,n;\bar{m},\bar{n})$ system. The lattice must now have $2N$ sites around a period. 
So, if we have a real 2-string (regarded as being in the lower half-plane), then we must have $N-1$ zeros in the upper half-plane and $N$ zeros 
in the lower half-plane.
Otherwise we must have $N$ zeros in each half-plane. 

In terms of the RSOS paths, we fix the last step of the path in the upper half-plane to be either $(2,3)$ or $(3,2)$ depending on the parity of $N$. 
This ensures that the half paths in the upper and lower half-planes both end at the ground state heights corresponding to the shaded band in Figure~
\ref{periodicPathsZeros}.
The contributing characters
are $\hat\chi_{h}^{(N-1)}(q)=\sum_E q^{E}$, as given by (\ref{eq:fincharbdry}), 
in the upper half plane and
\begin{equation}
\hat\chi_{h}^{(N)}(\bar{q})=\sum_E\bar{q}^{E}=
\sum_{\sigma}\bar{q}^{\sum_{j=1}^{N-1}(2N-j)
H(\sigma_{j-1},\sigma_{j},\sigma_{j+1})}
\end{equation}
in the lower half-plane where $\bar{q}$ denotes the complex conjugate of the modular nome $q$. Generalizing the arguments of Melzer~\cite{Melzer} to this non-unitary theory, the modular invariant partition function admits the finitization
\begin{equation}
Z^{(N)}(q)=\chi_0^{(N-1)}(q)\chi_0^{(N)}(\bar{q})+\chi_{-\frac{1}{5}}^{(N-1)}(q)\chi_{-\frac{1}{5}}^{(N)}(\bar{q})\to Z(q),\qquad N\to\infty
\end{equation}
The correct counting of states follows from the identity
\begin{equation}
Z^{(N)}(1)=F_{N-1}F_N+F_NF_{N+1}=F_{2N}
\end{equation}

In the $s=\bar{s}=1$ sector, the state with 1-strings, but with no 
short 2-strings corresponds to the vacuum state $|0\rangle=|0,0\rangle=|0\rangle\otimes|\bar{0}\rangle$.
The state with one short
2-string in the upper half-plane furthest from the real axis corresponds
to the state $L_{-2}|0\rangle$. Moving the short 2-string downwards through
the long 2-strings increases the level by 1 unit for each permutation, thus
creating the $L_{-n}|0\rangle$ state. The mirror image argument applies to the lower half-plane
to give states such as $\bar{L}_{-n}|\bar{0}\rangle$. 
In the $s=\bar{s}=2$ sector, the lowest energy state $|\Phi\rangle=|\!-\frac{1}{5},-\frac{1}{5}\rangle=|\!-\frac{1}{5}\rangle\otimes|\overline{-\frac{1}{5}}\rangle $ has no 
1-strings and no short 2-strings. 
The first excited state in the sector $V_{1}\otimes\bar{V}_{1}$ contains
one short 2-string on the top of all long 2-strings and correspond to
$L_{-1}|\Phi\rangle$. Every time we lower a short 1-string below
a long 2-string we obtain one extra unit of energy, hence we generate
all the $L_{-n}|\Phi\rangle$ and similarly for the lower half plane
with $\bar{L}_{-n}$ Virasoro modes. 
These correspondences between Virasoro descendants, RSOS paths and the
zero patterns can be read off from Figure~\ref{periodiccorrespondence}.

\subsubsection{Periodic {boundary condition} with a seam}

\begin{figure}[htbp]
\begin{centering}
\includegraphics[height=7cm]{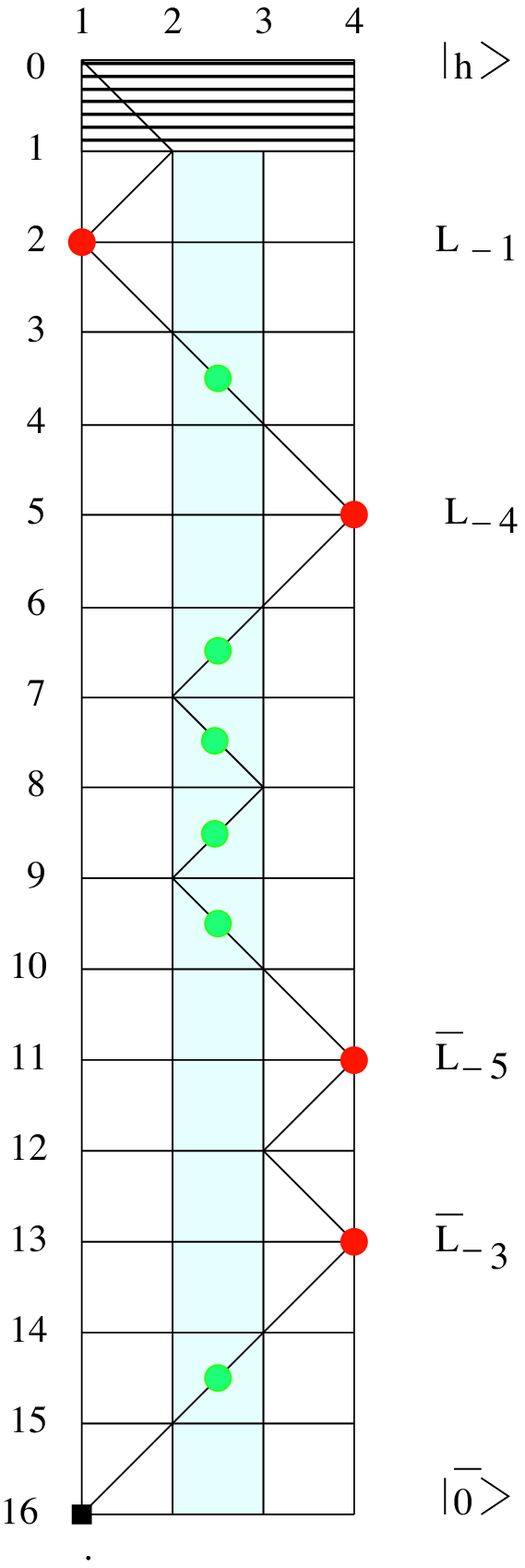}~~~\includegraphics[height=7cm]{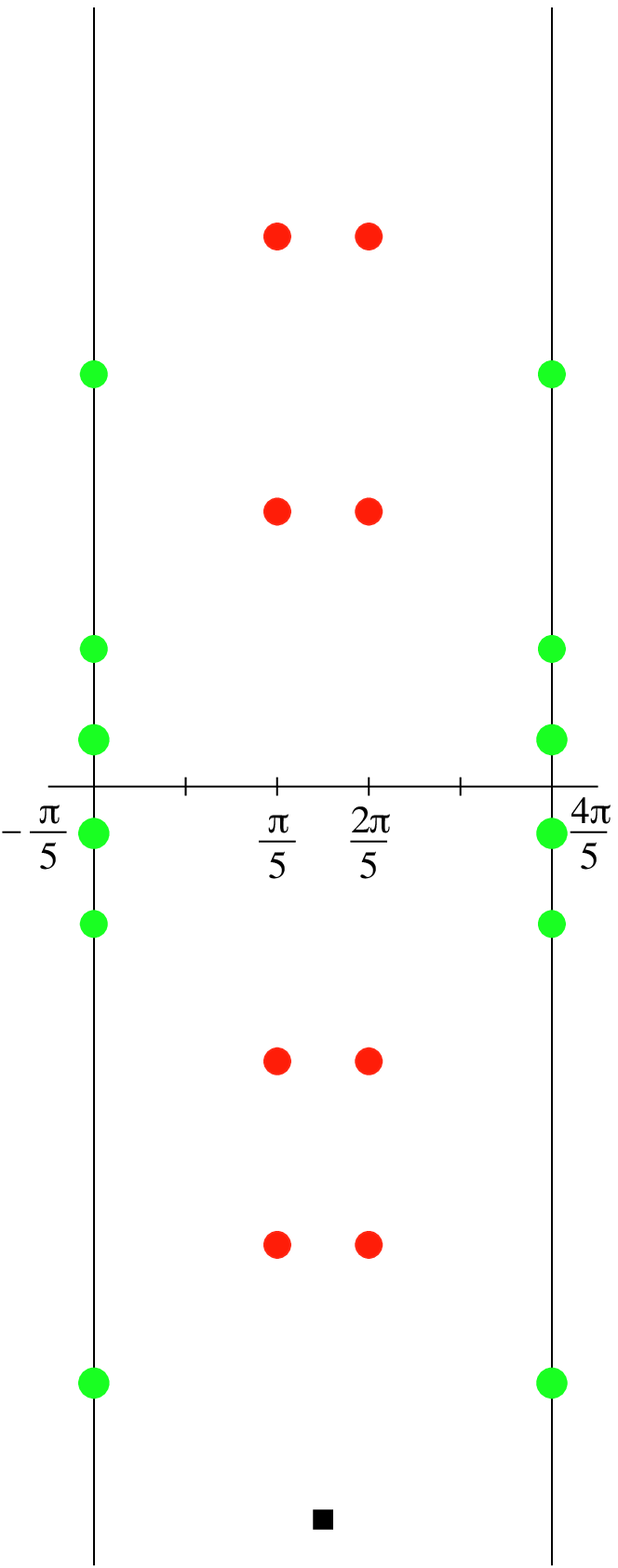}
~~~~~~~~~~~~\includegraphics[height=7cm]{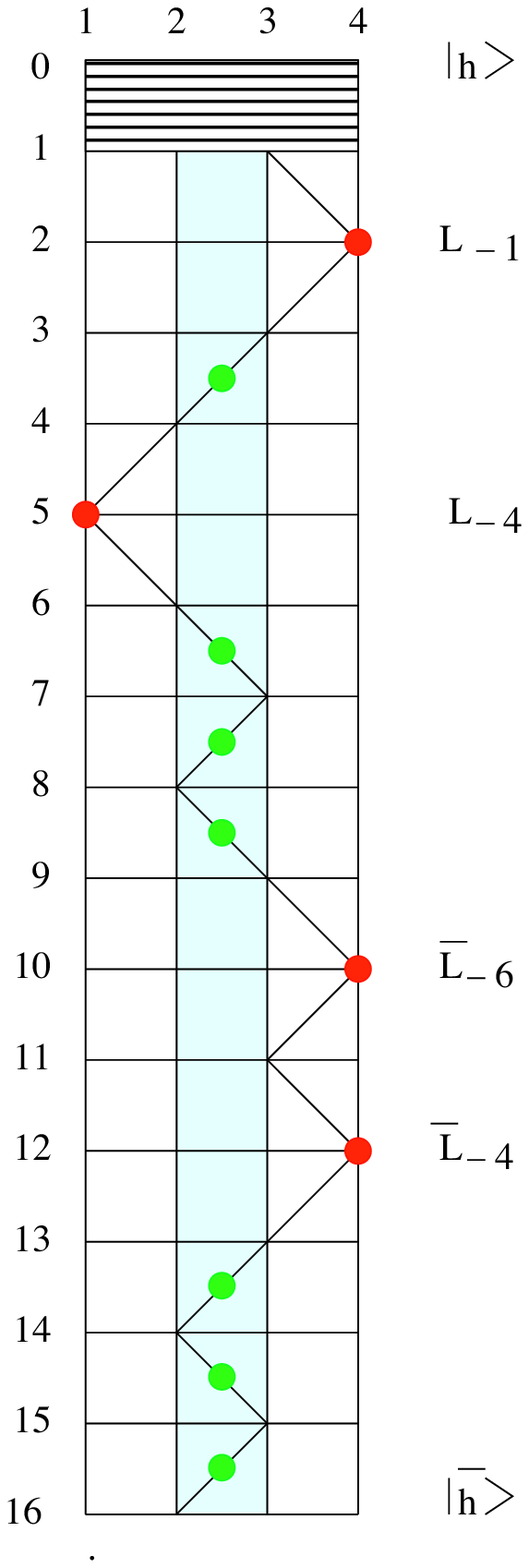}~~~\includegraphics[height=7cm]{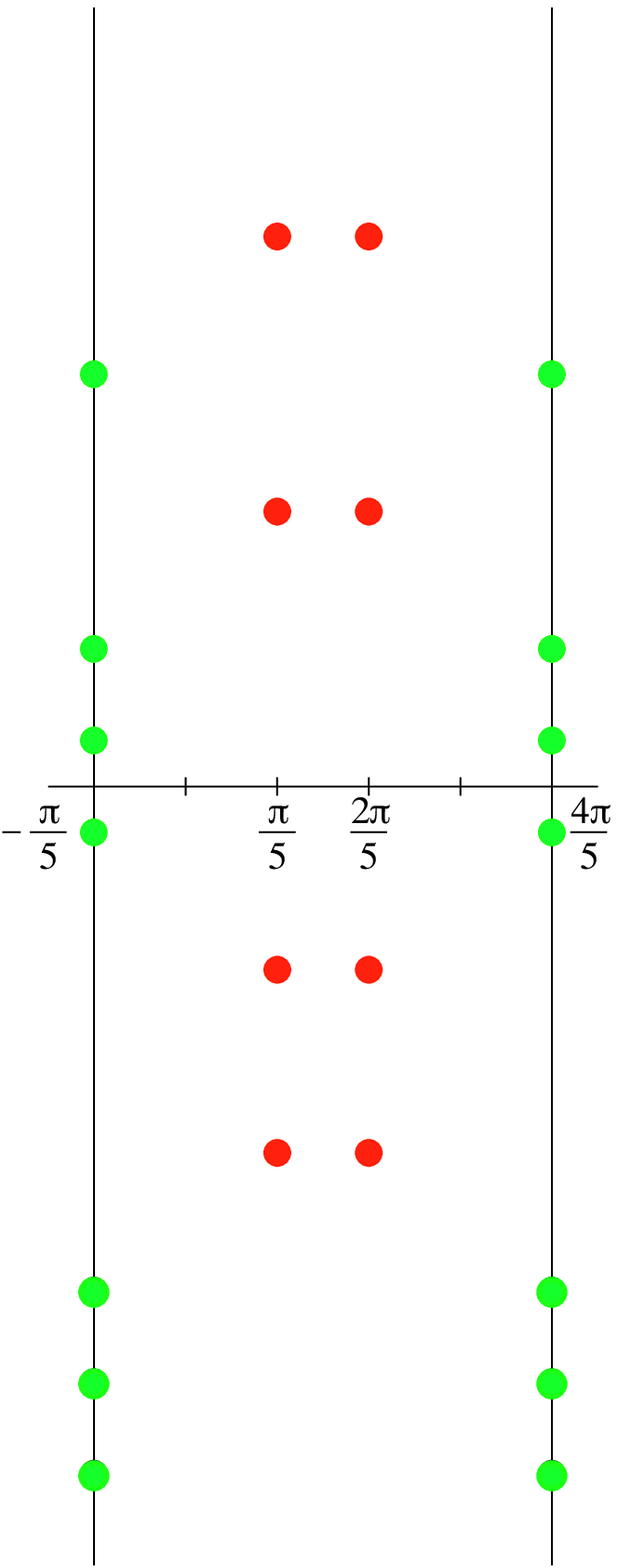}~
~~~~~~~~~~~\includegraphics[height=7cm]{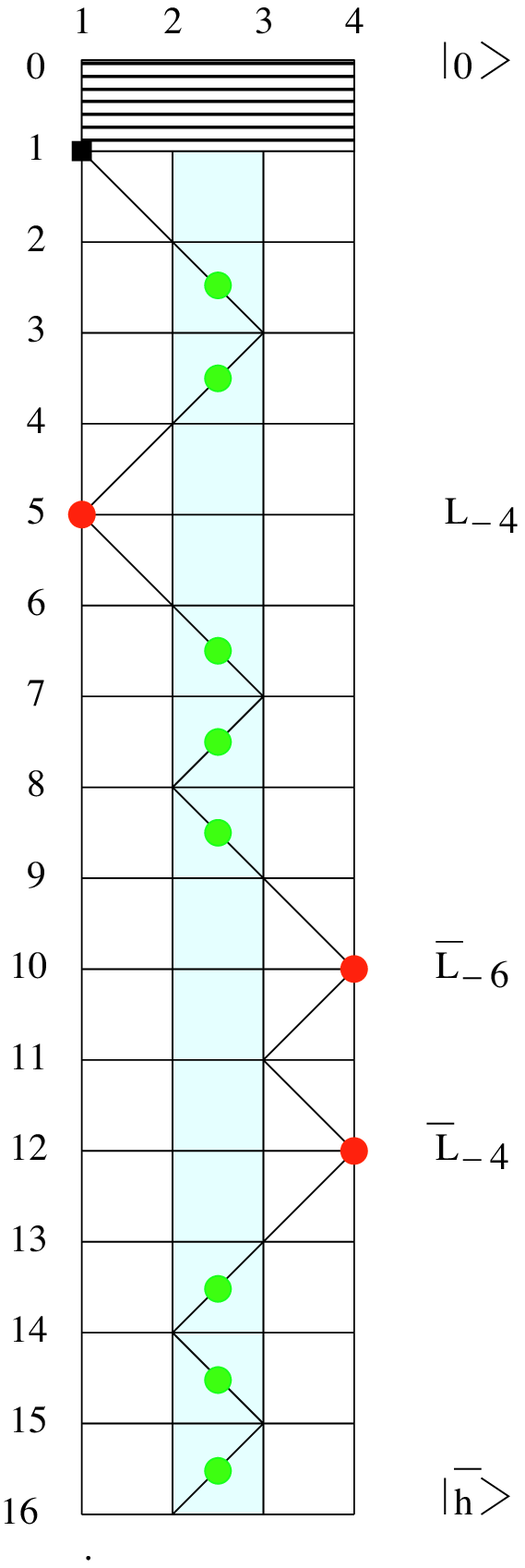}~~
~\includegraphics[height=7cm]{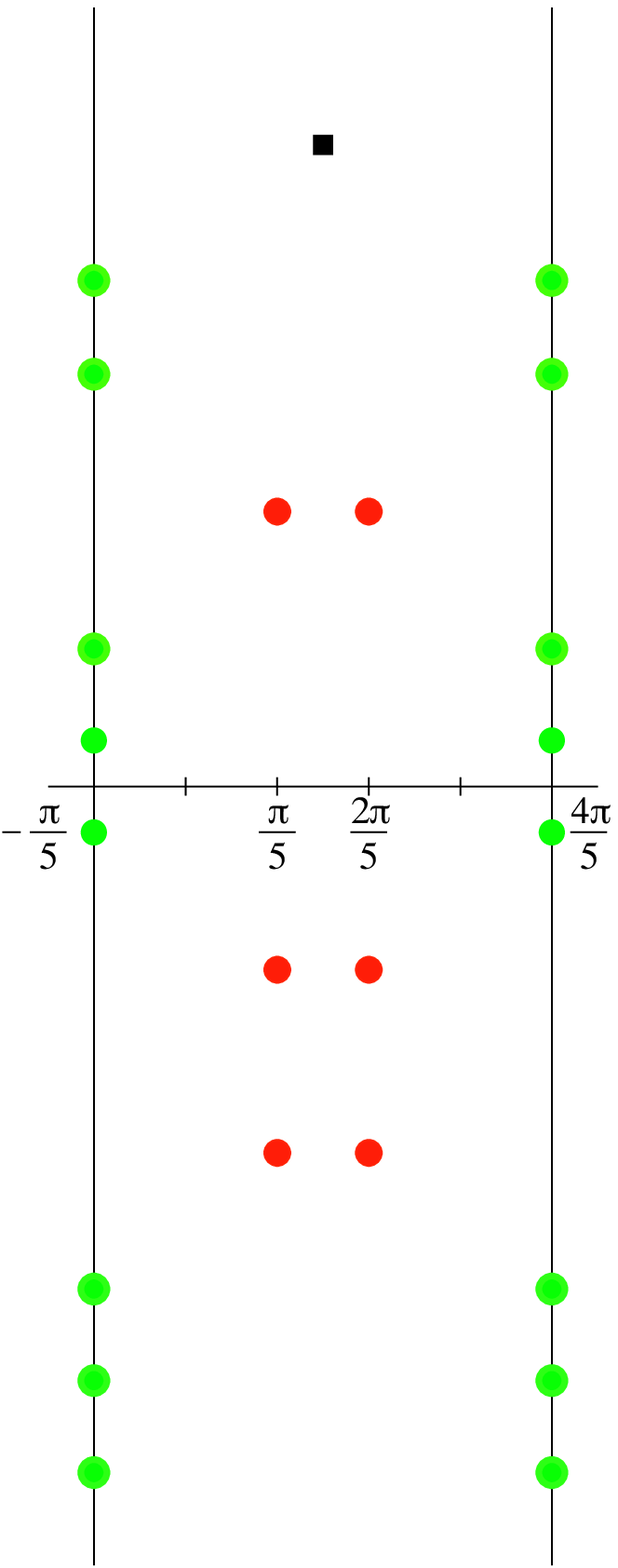}
\par\end{centering}

\vspace{-0.1in}
 \caption{Identification, for the seam transfer matrix eigenvalues, between the zero patterns, RSOS paths
and Virasoro modes in the case of an $(r,s)=(1,2)$ seam 
($\xi\to i\infty$). The heights across the seam differ by $\pm 1$. The RSOS paths are restricted to pass through the shaded ground-state band at the level of the real axis. The remaining RSOS paths are related to these by the $\mathbb{Z}_2$ $A_4$ symmetry.\label{seamcorrespondence}}
\end{figure}

Consider the periodic transfer matrix (\ref{seamT}) with a simple seam with parameter $\xi$. By varying $\xi $ from $0$ to $i\infty$, the two endpoints are described as 
follows:  for $\xi=0$ we have the $(r,s)=(1,1)$ identity seam which disappears, so we recover the results of the periodic boundary
condition analyzed in the previous subsection. In the $\xi\to i\infty$ limit, which gives the $(r,s)=(1,2)$ seam, we find the identification
between string patterns, RSOS paths and Virasoro modes summarized in Figure~\ref{seamcorrespondence}. 
Comparing with the periodic case we can see that, in the $\xi=i\infty$
limit, the sector selection rule is that a 1-string can appear either in the lower or in the upper
half plane but not in both. The identification is otherwise as in
the periodic case but the Hilbert space in the $N\to\infty$ limit is
\begin{equation}
\mathcal{H}=V_{0}\otimes\bar{V}_{-\frac{1}{5}}+V_{-\frac{1}{5}}\otimes\bar{V}_{0}+V_{-\frac{1}{5}}\otimes\bar{V}_{-\frac{1}{5}}
\end{equation}
corresponding to the twist partition function
\begin{equation}
Z_{\text{\tiny seam}}(q)=\chi_0(q)\chi_{-\frac{1}{5}}(\bar{q})+\chi_{-\frac{1}{5}}(q)\chi_{0}(\bar{q})+\chi_{-\frac{1}{5}}(q)\chi_{-\frac{1}{5}}(\bar{q})
\end{equation}
with the finitized version
\begin{equation}
Z_{\text{\tiny seam}}^{(N)}(q)=\chi_0^{(N-1)}(q)\chi_{-\frac{1}{5}}^{(N-1)}(\bar{q})+\chi_{-\frac{1}{5}}^{(N-1)}(q)\chi_{0}^{(N-1)}(\bar{q})+\chi_{-\frac{1}{5}}^{(N-1)}(q)\chi_{-\frac{1}{5}}^{(N-1)}(\bar{q})
\end{equation}
The correct counting of states follows from the identity
\begin{equation}
Z_{\text{\tiny seam}}^{(N)}(1)=F_{N-1}F_{N}+F_NF_{N-1}+F_N^2=F_{2N}
\end{equation}
The corresponding Virasoro highest weights states are denoted by $\vert{\phi}\rangle$,
$\vert\bar\phi\rangle$ and $\vert\Phi\rangle$ respectively. 
In the language of 1-d configurational sums, we see that the contributing part of the RSOS paths is not periodic. Instead, the heights across the seam differ by $\pm 1$ in accord with the weights of the $(1,2)$ seam (\ref{12seam}).

\paragraph{Seam flow.}
\begin{figure}[thbp]
\begin{centering}
\includegraphics[width=3.cm]{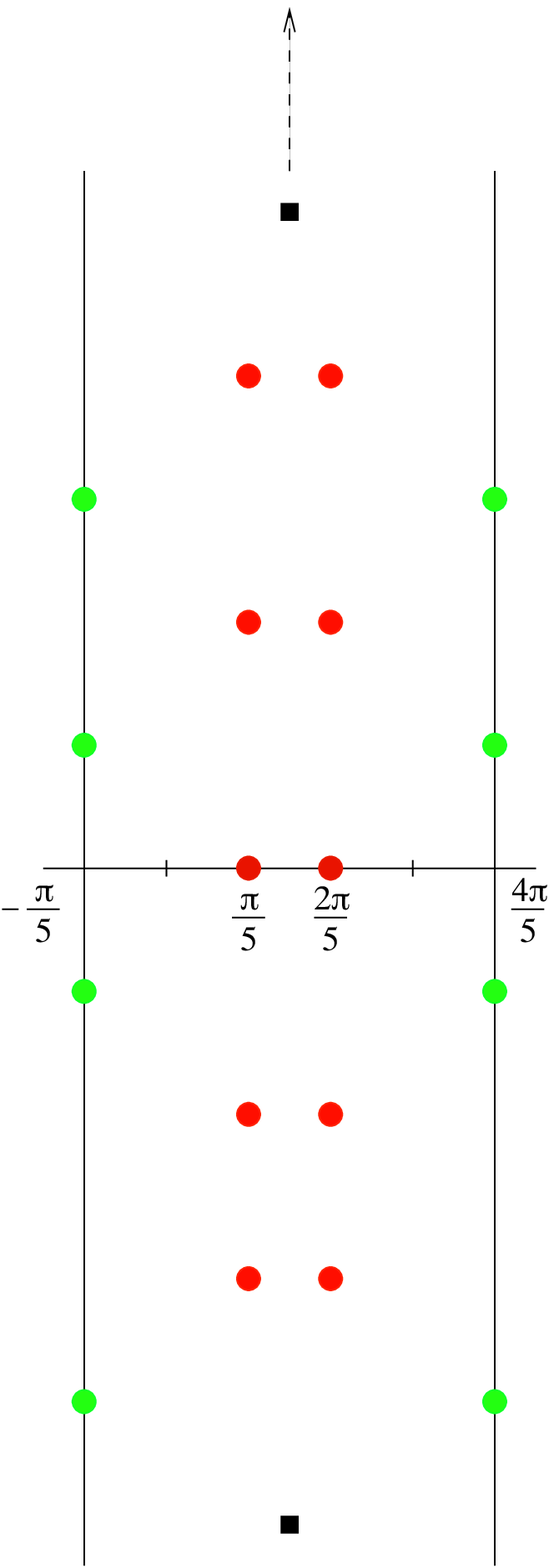}~~~~~~~~~~~~~~~~~\includegraphics[width=3.cm]{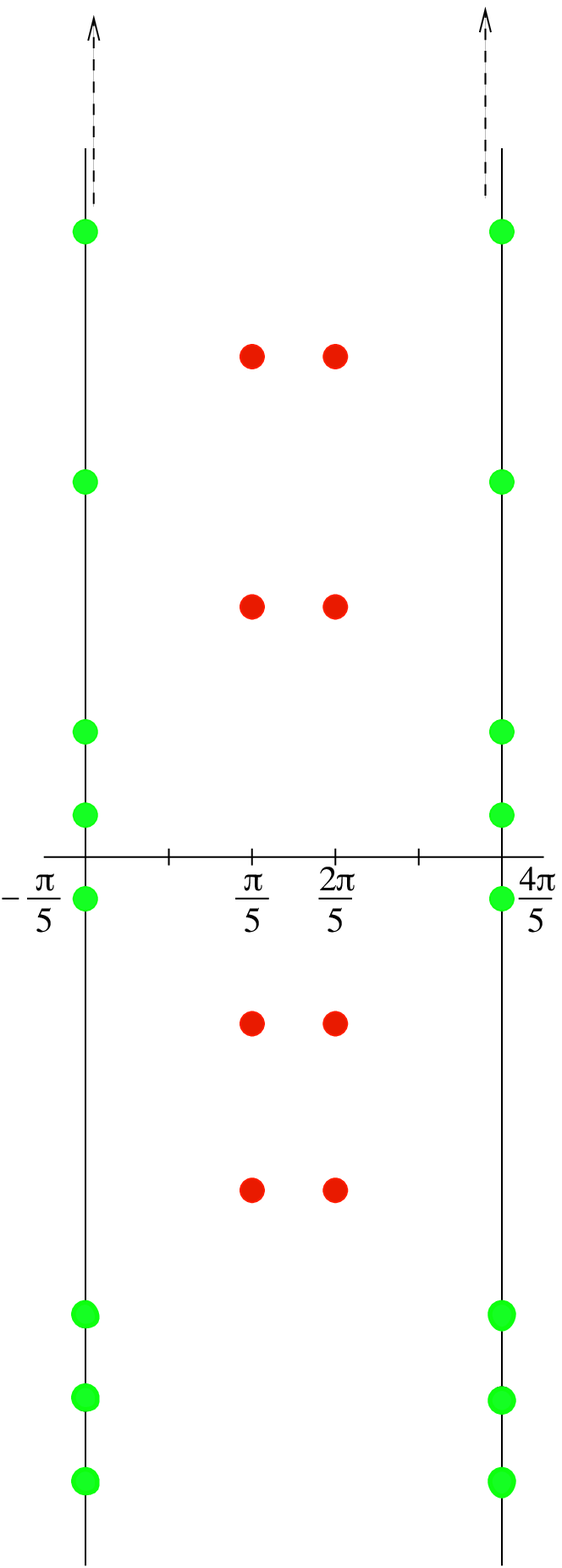}~
~~~~~~~~~~~\includegraphics[width=3.cm]{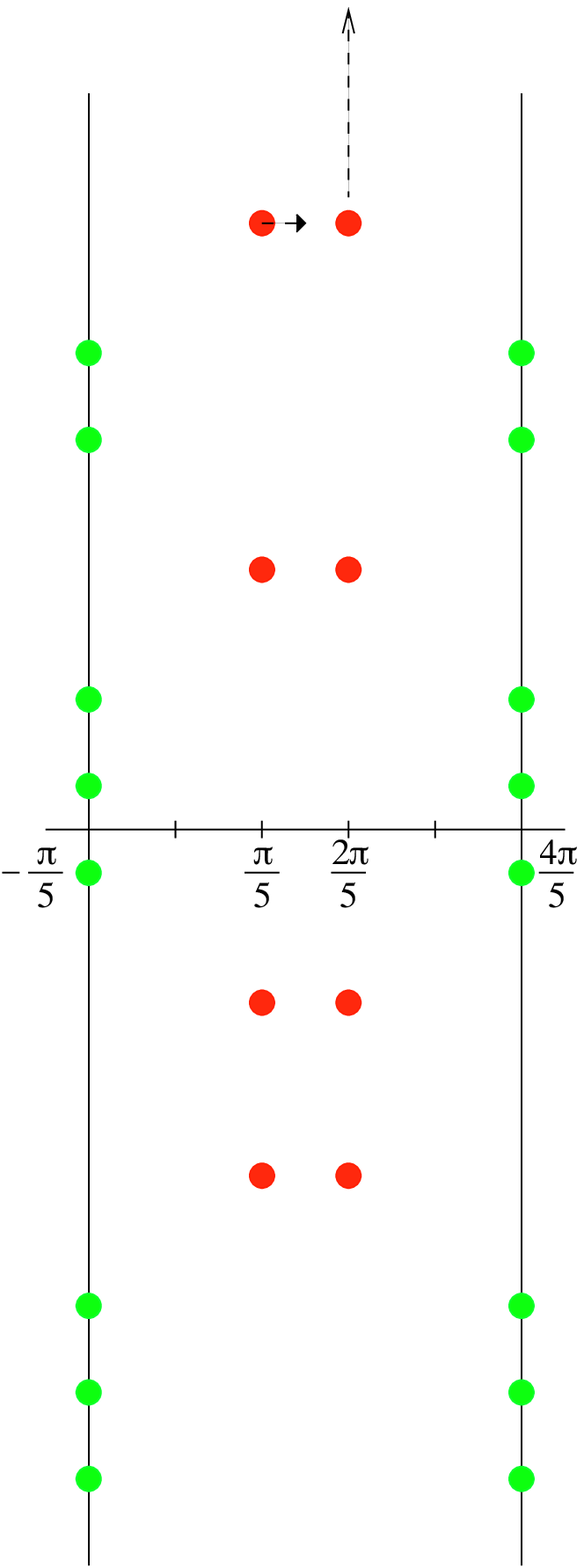}~~~
\par\end{centering}

\vspace{-0.1in}
 \caption{Depending on the outermost configurations the three different mechanisms A, B, C,
which appear in the flow $\xi\to i\infty$.\label{seamFlow}}
\end{figure}

We can analyse the flow in $\xi$ in terms of the various descriptions as indicated. 
In terms of the zeros, we identify three mechanisms depending on
the sector:
\begin{enumerate}
\item[A.] If the outermost string is a 1-string, it flows to infinity (in the same half-plane) as $\xi\to i\infty$. 
\item[B.] If the outermost string is a long 2-string, it flows to infinity (in the same half-plane) as $\xi\to i\infty$. 
\item[C.] If the outermost string is a short 2-string, one of the zeros flows
to infinity (in the same half-plane) and the other becomes a 1-string with real part $\frac{3\pi}{10}$. 
\end{enumerate}
These mechanisms, shown in the left, middle and right respectively in Figure~\ref{seamFlow}, are similar to the three mechanisms observed~\cite{FPR} for the boundary flow in the tricritical Ising model.

\goodbreak
In terms of Virasoro states, this is summarized as:
\begin{enumerate}
\item[1.] Due to type A flows with $|0\rangle\rightarrow|\bar{\phi}\rangle$: 
\[
L_{-N_{1}}...L_{-N_{n}}\bar{L}_{-\bar{N}_{1}}....\bar{L}_{-\bar{N}_{n}}|0
\rangle\rightarrow L_{-N_{1}}...L_{-N_{n}}\bar{L}_{-\bar{N}_{1}+1}..
..\bar{L}_{-\bar{N}_{n}+1}|\bar{\phi}\rangle
\]
\item[2.] Due to type B flows with $|\Phi\rangle\rightarrow|{\Phi}\rangle$: 
\[
L_{-N_{1}}...L_{-N_{n}}\bar{L}_{-\bar{N}_{1}}....\bar{L}_{-\bar{N}_{n}}|
\Phi\rangle\rightarrow L_{-N_{1}}...L_{-N_{n}}\bar{L}_{-\bar{N}_{1}+1}..
.\bar{L}_{-\bar{N}_{n}+1}|\Phi\rangle
\]
\item[3.] Due to type C flows with $|\Phi\rangle\rightarrow|{\phi}\rangle$: 
\[
(L_{-N_{1}}...L_{-N_{n}}\bar{L}_{-\bar{N}_{1}}....\bar{L}_{-\bar{N}_{n}})
\bar{L}_{-1}|\Phi\rangle\rightarrow(L_{-N_{1}}...L_{-N_{n}}\bar{L}_{-\bar{N}_{1}+1}.
...\bar{L}_{-\bar{N}_{n}+1})|\phi\rangle
\]
\end{enumerate}
The three mechanisms for the flow of the patterns of zeros are confirmed numerically.
It follows that the flow, under the three mechanisms, of the first few states of the identity defect as in Figure~\ref{seamFlow} is as shown in Table~\ref{seamFlowTab}.
\begin{table}[thb]
\begin{center}
\begin{tabular}{|c|c|c|c|}
\hline 
Level & Trivial Defect & Non-trivial Defect & Level\tabularnewline
\hline 
h.w. state & $|0\rangle$ & |$\bar{\phi}\rangle$ & h.w. state\tabularnewline
h.w. state & |$\Phi\rangle$ & |$\Phi\rangle$ & h.w. state\tabularnewline
\cline{1-2} 
1 & $\bar{L}_{-1}|\Phi\rangle$ & $|\phi\rangle$ & h.w. state\tabularnewline
\cline{3-4} 
1 & $L_{-1}|\Phi\rangle$ & $L_{-1}|\Phi\rangle$ & 1\tabularnewline
\cline{1-2} 
2 & $\bar{L}_{-2}|0\rangle$ & $\bar{L}_{-1}$$|\bar{\phi}\rangle$ & 1\tabularnewline
2 & $\bar{L}_{-2}|\Phi\rangle$ & $\bar{L}_{-1}|\Phi\rangle$ & 1\tabularnewline
2 & $\bar{L}_{-1}L_{-1}|\Phi\rangle$ & $L_{-1}|\phi\rangle$ & 1\tabularnewline
\cline{3-4} 
2 & $L_{-2}|\Phi\rangle$ & $L_{-2}|\Phi\rangle$ & 2\tabularnewline
2 & $L_{-2}|0\rangle$ & $L_{-2}|\bar{\phi}\rangle$ & 2\tabularnewline
\cline{1-2} 
3 & $\bar{L}_{-3}|0\rangle$ & $\bar{L}_{-2}$$|\bar{\phi}\rangle$ & 2\tabularnewline
3 & $\bar{L}_{-3}|\Phi\rangle$ & $\bar{L}_{-2}|\Phi\rangle$ & 2\tabularnewline
3 & $\bar{L}_{-2}L_{-1}|\Phi\rangle$ & $\bar{L}_{-1}L_{-1}|\Phi\rangle$ & 2\tabularnewline
3 & $\bar{L}_{-1}L_{-2}|\Phi\rangle$ & $L_{-2}|\phi\rangle$ & 2\tabularnewline
\cline{3-4} 
3 & $L_{-3}|\Phi\rangle$ & $L_{-3}|\Phi\rangle$ & 3\tabularnewline
3 & $L_{-3}|0\rangle$ & $L_{-3}|0\rangle$ & 3\tabularnewline
\hline 
4 & $\bar{L}_{-1}\bar{L}_{-3}|\Phi\rangle$ & $\bar{L}_{-2}|\phi\rangle$ & 2\tabularnewline
\hline 
\end{tabular}
\par\end{center}
\caption{The flow from each state in the trivial identity defect
Hilbert space to its corresponding state in the non-trivial defect Hilbert space up
to the second order descendent level in the defect Hilbert space.
\label{seamFlowTab}}
\end{table}

As expected from defect CFT, the flow is 
\begin{equation}
V_{0}\otimes\bar{V}_{0}+V_{-\frac{1}{5}}\otimes\bar{V}_{-\frac{1}{5}}
\ \mapsto\ V_{0}\otimes\bar{V}_{-\frac{1}{5}}+V_{-\frac{1}{5}}\otimes\bar{V}_{0}+V_{-\frac{1}{5}}\otimes\bar{V}_{-\frac{1}{5}}
\end{equation}
Taking instead the limit $\xi\rightarrow-i\infty$ gives a similar result 
but with $\phi$ and $\bar{\phi}$ interchanged and $L_{-n}$
being the operator augmented to $L_{-n+1}$ instead of the $\bar{L}_{-n}$ modes.
The corresponding physical flows were confirmed level by level using defect TCSA in \cite{Bajnok:2013waa}.

\subsection{Continuum scaling limit in the critical case}

\begin{figure}[thbp]
\begin{centering}
\includegraphics[width=3.5cm]{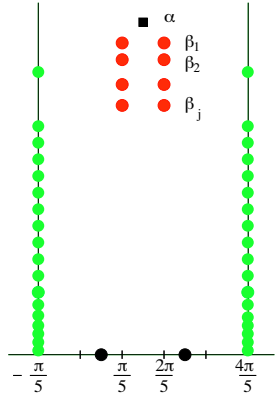}
\par\end{centering}
\vspace{-.1in}
\caption{Pattern of zeros of a finite excitation for 
large $N$ showing the scaling region a distance $\frac{3}{5}\log N$ out from the real axis.\label{largeNbdy}}
\end{figure}

In this section, for the critical case, we explain how the distribution of zeros scale in the continuum
scaling limit for finite energy states. As $N\to\infty$ the spacing of the zeros 
becomes more dense. Quantitatively, the imaginary part of the outer most
zeros grow as $\frac{3}{5}\log N$ so the spacing between zeros shrinks to $0$ as $\frac{3}{5}\frac{\log N}{N}$. 
The pattern of zeros for a finite energy state is
shown schematically in Figure~\ref{largeNbdy}. The imaginary part of the 1-string
is denoted by $\alpha$, while the imaginary part of the short 2-strings
are denoted by $\beta_{j}$. There are a finite number of short 2-string excitations.
The $\alpha$ and $\beta_{j}$ variables together with the imaginary
parts of the long 2-strings furthest from the real axis all scale as $\frac{3}{5}\log N$ in
the continuum scaling limit. These configurations correspond to Virasoro descendants
with a finite number of Virasoro modes and to paths which, except over a
finite interval, alternate between heights $2$ and $3$. Similar 
observations apply for periodic boundary conditions, with or without a seam, except
that the  patterns in the upper and lower half of the plane need not be the same. In the continuum scaling limit, the
zeros in the scaling regions in the upper and lower half-planes  are effectively 
``infinitely far'' from each other and thus independent of each other.

\subsection{Classification of states in the off-critical theory}

\begin{figure}[thbp]
\begin{centering}
\includegraphics[height=7.5cm]{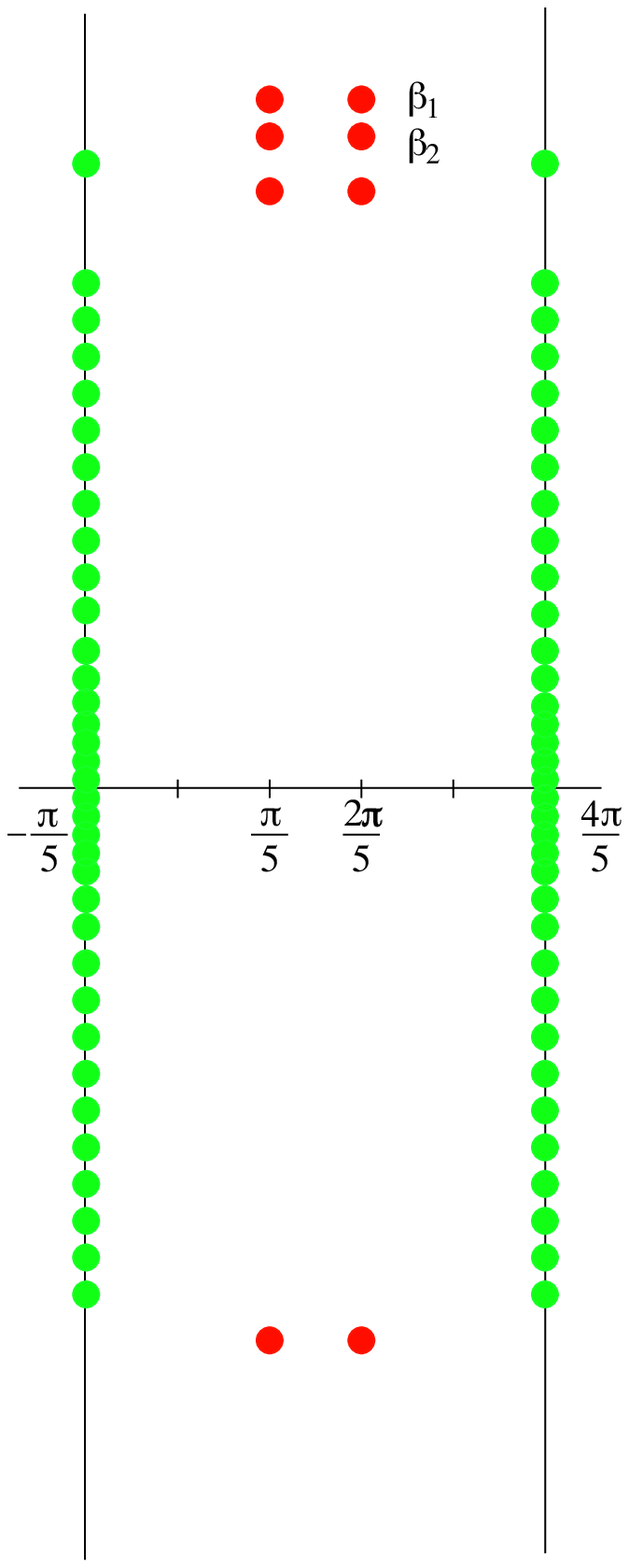}~~~
~~~\includegraphics[height=7.5cm]{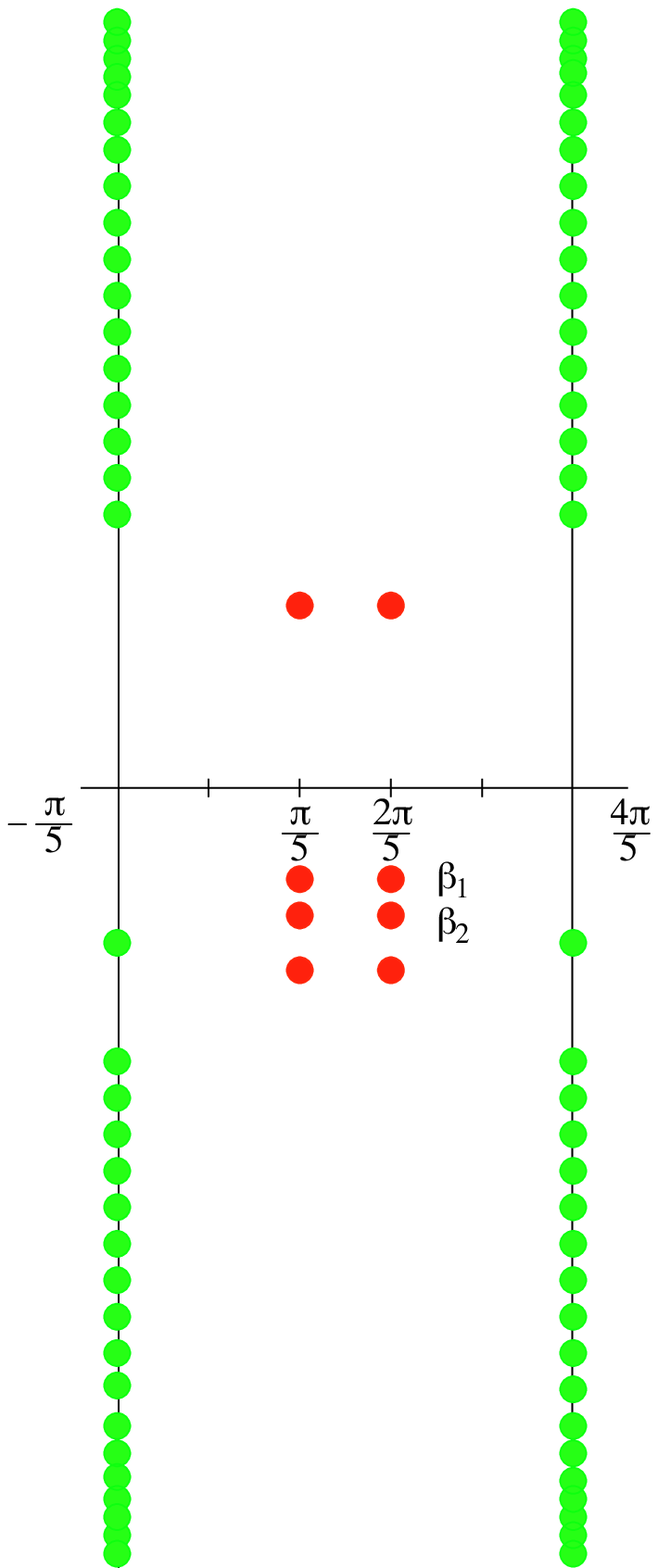}
\par\end{centering}
\vspace{-.1in}
\caption{Periodicity of the off-critical transfer matrix brings the scaling regions in the upper and lower half-planes close to each other across the boundary of the period rectangle. On the left, the massive zeros are shown using the variable $u$, while on the right, the shifted variable $u+\frac{1}{2}\log q$ is used.
\label{periodRect}}
\end{figure}

In the off-critical theory, the Boltzmann weights are given in terms
of the elliptic function $s(u)=\vartheta_{1}(u,q)$. Unlike the sine function, this elliptic function 
is quasi-periodic in the imaginary direction
\begin{equation}
\vartheta_{1}(u+\pi,q)=-\vartheta_{1}(u,q),\qquad
\vartheta_{1}(u-i\log q,q)=-q^{-1}e^{-2iu}\vartheta_{1}(u,q)
\end{equation}
It follows that the eigenvalues of the transfer matrices in the various geometries 
are entire functions of $u$ in the period rectangle 
\begin{equation}
\mbox{period rectangle}=\{\Re u\in[-\tfrac{\pi}{5},\tfrac{4\pi}{5}],\ 
\Im u\in[-\tfrac{1}{2}\log q,\tfrac{1}{2}\log q]\}
\end{equation}
We can therefore restrict our analysis to this period rectangle. Moreover, within this period rectangle,
the eigenvalues are completely characterized by their zeros. 
Actually the renormalized transfer matrices are periodic in this period rectangle. 

We are interested in the massive Lee-Yang model which is the continuum scaling limit of the 
off-critical lattice Lee-Yang model. The parameter $q^2$ measures the departure from criticality. 
As we will see, we have to tune
the square of the elliptic nome as $q^2\propto N^{-5/3}$ as $N\to\infty$ for the continuum scaling limit to exist. Numerically, for
reasonable sizes of $N$, we find that the locations of the zeros in the complex $u$-plane follow the same patterns found in the critical case. The observed deviations from the critical locations
are exponentially small in $N$. 
What is conceptually different is that, while in the critical case
the upper and lower scaling regions are ``infinitely far'' apart,
in the massive case periodicity glues them together as shown in Figure~\ref{periodRect}.

\section{Critical TBA Equations}

In this section we exploit the analytic structure of the transfer
matrix eigenvalues to turn the functional relation (\ref{eq:funcrel})
into TBA integral equations. We start by reviewing the critical case with periodic boundary conditions before adding a seam and moving 
to the massive case. In the periodic critical case, the central charge and conformal dimension of the single non-trivial primary 
field were calculated analytically in \cite{BLZ} following the methods of \cite{KlumP91,KlumP92}.

\subsection{Critical TBA}

The critical TBA equations are derived by solving the functional relation
\begin{equation}
t(u)t(u+\lambda)=1+t(u+3\lambda)
\end{equation}
taking into account the analytic structure of the function $t(u)$.
The function $t(u)$ is factorized  according to its large volume behaviour
as 
\begin{equation}
t(u)=f(u)g(u)l(u)\label{eq:fgl}
\end{equation}
where $\log f(u)=O(N)$, $\log g(u)=O(1)$ and $\log l(u)=O(N^{-1})$.
The leading order term satisfies
\begin{equation}
f(u)f(u+\lambda)=1\label{eq:ff1}
\end{equation}
 and accounts for the order $N$ zeros and poles of the normalization.
The function $g(u)$ also satisfies
\begin{equation}
g(u)g(u+\lambda)=1\label{eq:gg1}
\end{equation}
and accounts for the order $1$ boundary- and seam-dependent zeros and poles.
An integral equation is derived for the remaining finite-size function $l(u)$.

\subsubsection{Periodic case}

The energy of the states is extracted from the finite-size corrections of the transfer matrix eigenvalues
\begin{equation}
\log t(u)=\log f(u)+\log g(u)-\frac{i}{N}(e^{-\frac{5i}{3}u}E^{+}-
e^{\frac{5i}{3}u}E^{-})\label{eq:criticalenergy}
\end{equation}
For physical values of $u$ with $0<u<{3\pi\over 5}$, the first term in the ${1\over N}$ correction 
dominates when $u\to +i\infty$ and the second term when $u\to -i\infty$.
The finite-size conformal Lee-Yang energies are 
\begin{equation}
E_{LY}=e^x E^{+}+e^{-x} E^{-},\qquad u=\frac{\lambda}{2}+\frac{3ix}{5}
\label{confELY}
\end{equation} 
These energies can be calculated analytically following the methods of \cite{KlumP91,KlumP92}.

As the analytic structure of the two sectors are different, we start
with the simpler $(r,s)=(1,2)$ or $\vert\Phi\rangle$ sector, which contains the lowest energy state or ``true vacuum''. 
For periodic boundary conditions, the number of faces $N$ is even. 
Using the periodicity of the transfer matrix $t(u)=t(u+\pi)$, after
an appropriate shift, we obtain
\begin{equation}
t(u-\frac{\pi}{5})t(u+\frac{\pi}{5})=1+t(u)
\end{equation}
The normalization (\ref{eq:pernorm}) introduces order $N$ zeros
at $-\frac{\pi}{5},\frac{4\pi}{5}$ and order $N$ poles at $\frac{\pi}{5},\frac{2\pi}{5}$.
We remove them by normalizing with the function $f(u)$ which satisfies
\begin{equation}
f(u-\frac{\pi}{5})f(u+\frac{\pi}{5})=f(u)
\end{equation}
The solution compatible with the analytic structure is 
\begin{equation}
f(u)=f_1(u)^N=
\Big(\frac{\sin\frac{5u}{3}+\sin
\frac{\pi}{3}}{\sin\frac{5u}{3}-
\sin\frac{\pi}{3}}\Big)^{N}=\Big(\!-\frac{\sin(\frac{5u}{6}+
\frac{\pi}{6})\sin(\frac{5u}{6}+\frac{2\pi}{6})}{\cos(\frac{5u}{6}+
\frac{\pi}{6})\cos(\frac{5u}{6}+\frac{2\pi}{6})}\Big)^{N}
\end{equation}
Observe that this solution also satisfies $f(u)f(u+\lambda)=1$ as
anticipated in (\ref{eq:ff1}). In the periodic case, there is no order $1$ boundary contribution so 
$g(u)=1$. 

In the large $N$ limit, the imaginary part of the position of the long 2-string furthest from the real $u$ axis approaches infinity
as $\frac{3}{5}\log \kappa N$ with 
\begin{equation}
\kappa=4\sin\frac{\pi}{3}=2\sqrt{3}
\end{equation}
This motivates introducing the real variable $x$ as a scaled vertical coordinate along the centre of the analyticity strip
\begin{equation}
u=\frac{3\pi}{10}+\frac{3ix}{5}\label{eq:uxrel}
\end{equation}
In terms of $x$, the functional equation takes the form 
\begin{equation}
t(x-i\frac{\pi}{3})t(x+i\frac{\pi}{3})=1+t(x)\label{eq:t(x)}
\end{equation}
In this variable 
\begin{equation}
f(x)=\left(\frac{\cosh x+\sin\frac{\pi}{3}}{\cosh x-\sin\frac{\pi}{3}}\right)^{N}
\end{equation}

\paragraph{Vacuum state.}
Focusing on the ground state, we divide (\ref{eq:t(x)})
by $t(x)$ and use the properties of the function $f(x)$ to obtain the standard scaling Lee-Yang $Y$-system 
\begin{equation}
\frac{t(x-i\frac{\pi}{3})t(x+i\frac{\pi}{3})}{t(x)}=\frac{l(x-i\frac{\pi}{3})
l(x+i\frac{\pi}{3})}{l(x)}=1+{1\over t(x)}\label{eq:tt/t}
\end{equation}
Due to our construction $l(x)$ is Analytic and Non-Zero in the analyticity strip $\Im x\in (-\frac{5\pi}{6},\frac{5\pi}{6})$ 
and its logarithm has Constant asymptotic (ANZC) as $x\to\pm\infty$. Following \cite{KlumP91,KlumP92}, we can thus take the logarithm
and solve the equation using Fourier transforms of the derivatives $[\log l(x)]'$ and so on. Evaluating the constants introduced by integrating gives
\begin{equation}
\log l(x)=-\varphi\star\log\Big(1+{1\over t(x)}\Big),
\quad \varphi(x)=-{4\sqrt{3}\cosh x\over 1+2\cosh 2x},
\quad\hat{\varphi}=\frac{1}{1-e^{k\frac{\pi}{3}}-e^{-k\frac{\pi}{3}}}
\end{equation}
where  the convolution is defined by
\begin{equation}
(f\star g)(x)=(g\star f)(x)={1\over 2\pi}\int_{-\infty}^\infty f(x-y)g(y) dy
\end{equation}
Observe that the kernel $\varphi(x)$
is related to the known~\cite{CardyMuss89} two-particle $S$-matrix (\ref{Smatrix}) of the Lee-Yang model as
$\varphi=-i\partial_x \log S(x)$.

We thus obtain the critical TBA equation on the lattice for the ground state with periodic boundary conditions
\begin{equation}
\log t(x)=\log f(x)-\varphi\star\log\Big(1+{1\over t(x)}\Big)
\end{equation}

We are interested in the continuum scaling limit with $N\to\infty$.
Analysing the scaling limit of the function $f(x)$, one can see that nontrivial
behaviours occur in the two scaling regions $x\sim\pm\log \kappa N$. 
It is thus natural to introduce two scaling functions  
\begin{equation}
e^{\epsilon^{\pm}(x)}=\lim_{N\to\infty}t(x\pm\log\kappa N)
\end{equation}
which relate to the left and right chiral halves of the theory. 
Using the explicit form of the source term $f(x)$, 
we obtain the behavior  in the two scaling regions 
\begin{equation}
\lim_{N\to\infty}\log f(x\pm\log\kappa N)=\lim_{N\to\infty} N
\log(1+e^{\mp x}\frac{1}{N})=e^{\mp x}
\end{equation}
which leads to the massless TBA equations
\begin{equation}
\epsilon^{\pm}(x)=e^{\mp x}- \varphi\star\log(1+e^{-\epsilon^{\pm}(x)})
\end{equation}
As the ground state is symmetric under $x\leftrightarrow-x$, we see that $\epsilon^+(x)=\epsilon^-(x)=\epsilon(x)$ for the ground state. Although this symmetry holds for spinless primary states, it is not generally true for excited states. 

\paragraph{Energies.}
The two scaling regions contribute to the energy of a general state through $E^{\pm}$,
which can be extracted from $\log l(x)$ using (\ref{eq:criticalenergy}) and (\ref{confELY})
\begin{eqnarray}
\frac{1}{N}\,(e^{x}E^{+}+e^{-x}E^{-})&=&-\log l(x)  =  \int_{-\infty}^{\infty}\frac{dy}{2\pi}\,
\varphi(x-y)\log\Big(1+{1\over t(y)}\Big)\nonumber \\
 & = & \left 
 \{ \int_{0}^{\infty}
 +\int_{-\infty}^{0}  \right \} \frac{dy}{2\pi}\,
\varphi(x-y)\log(1+e^{-\epsilon(y)})
\end{eqnarray}
We focus first on the $E^{+}$ contribution coming from the $x\to+\infty$ scaling region and the integral
$\int_{0}^{\infty}$. By shifting the integrals $y\mapsto y+\log \kappa N$ we
can write
\begin{equation}
e^{x}E^{+}=N\int_{-\log\kappa N}^{\infty}\frac{dy}{2\pi}\,
\varphi(x-y-\log\kappa N)\log(1+e^{-\epsilon(y+\log \kappa N)})
\end{equation}
In the $N\to\infty$ limit 
\begin{equation}
\lim_{N\to\infty}N\varphi(x-\log\kappa N)=-\lim_{N\to\infty}
\frac{\kappa N\cosh(x-\log\kappa N)}{\cosh(2x-2\log\kappa N)-
\cos\frac{2\pi}{3}}=-e^{x}
\end{equation}
which leads to
\begin{equation}
E^{+}=-\int_{-\infty}^{+\infty}\frac{dy}{2\pi}\,e^{-y}
\log(1+e^{-\epsilon^{+}(y)})\label{eq:E-}
\end{equation}
A similar calculation based on the integral $\int_{-\infty}^{0}$
gives the result 
\begin{equation}
E^{-}=-\int_{-\infty}^{+\infty}\frac{dy}{2\pi}\,e^{y}
\log(1+e^{-\epsilon^{-}(y)})\label{eq:E+}
\end{equation}

\paragraph{Central charge.}
Consider the ground state TBA with $\epsilon^+(x)=\epsilon^-(x)=\epsilon(x)$ and set
\begin{equation}
a(x)={1\over t(x)}=e^{-\epsilon(x)},\quad A(x)=1+a(x),\quad \ell a=\log a(x),\quad \ell A=\log A(x)
\end{equation}
Then differentiating the TBA gives
\begin{eqnarray}
-\ell a(x)=e^{-x}- \varphi\star\ell A,\qquad -\ell a'(x)=-e^{-x}-\varphi\star\ell A'
\end{eqnarray}
The solution of this TBA has flat plateaus in the asymptotic regions as $x\to\pm\infty$. 
The asymptotic values are
\begin{equation}
a(\infty)={2\over \sqrt{5}+1}={\sqrt{5}-1\over 2},\qquad a(-\infty)=0
\end{equation}
Indeed, using the fact that
\begin{equation}
\int_{-\infty}^\infty \!\frac{dy}{2\pi}\,\varphi(y)=-1
\end{equation}
it is easily checked that the asymptotic value of $a=a(+\infty)$ must be a solution of ${1\over a}=1+a$.

Next, using the fact that the kernel is even $\varphi(x)=\varphi(-x)$, integrating by parts and changing the integration variable to $a$, it follows that
\begin{eqnarray}
&&\int_{-\infty}^\infty \!\!dy\,[\ell a'(y)\ell A(y)-\ell a(y) \ell A'(y)]=\int_{-\infty}^\infty \!\!dy\,e^{-y}[\ell A(y)-\ell A'(y)]=2\int_{-\infty}^\infty \!\!dy\,e^{-y}\ell A(y)=\quad\nonumber\\
&&   -  4\pi E^+
=\int_{a(-\infty)}^{a(\infty)} 
da\Big[{\log(1+a)\over a}-{\log a\over 1+a}\Big]=2L_+\Big({\sqrt{5}-1\over 2}\Big)
=2L\Big({3-\sqrt{5}\over 2}\Big)={2\pi^2\over 15}
\end{eqnarray}
where the Rogers dilogarithm functions are defined by
\begin{align}
L(x)&=-{1\over 2} \int_0^x \!\!da\Big[{\log(1-a)\over a}+{\log a\over 1-a}\Big]\\
L_+(x)&={1\over 2}\int_0^x \!\!da\Big[{\log(1+a)\over a}-{\log a\over 1+a}\Big]=L\Big({x\over 1+x}\Big)
\end{align}
This gives the value of the energy integrals $E^+=E^-$. In particular, at the isotropic point 
with $u={\lambda\over 2}={3\pi \over 10}$ and $x=0$, we find the effective central charge
\begin{equation}
E_{LY}=E^++E^-=2E^+=-{\pi\over 15}=-{\pi c_\text{\tiny eff}\over 6},\qquad c_\text{\tiny eff}={2\over 5}
\end{equation}
Hence the central charge is
\begin{equation}
c=c_\text{\tiny eff}+24h_\text{\tiny min}={2\over 5}-{24\over 5}=-{22\over 5},\qquad h_\text{\tiny min}=-{1\over 5}
\end{equation}

\paragraph{Excited states.}
The eigenvalues of the transfer matrix are characterized by their patterns of zeros in the analyticity strip $-{\pi\over 5}<\Re u<{4\pi\over 5}$. For all eigenvalues, long 2-strings occur at the boundaries $\Re u=-{\pi\over 5},{4\pi\over 5}$ of this analyticity strip. In the thermodynamic limit $N\to\infty$, these 2-strings become dense defining the boundaries of the analyticity strip.
For finite excitations above the ground state, additional short 2-strings can occur at 
\begin{equation}
u_j=\begin{cases}
\frac{\pi}{5}+i\beta_{j}\\
\frac{2\pi}{5}+i\beta_{j}
\end{cases}
\end{equation}
Depending on the sector, single 1-strings can also occur furthest out from the real $u$ axis in the upper- and lower-half $u$-plane at
\begin{equation}
u_0=\frac{3\pi}{10}+i\alpha
\end{equation}
In the scaling regions, at the edge of the distributions of zeros, in the upper/lower half $u$-plane respectively, the zeros approach infinity in the thermodynamic limit as 
\begin{equation}
\alpha=\frac{3}{5}(\pm\log\kappa N+\tilde{\alpha}^{\mp}),\qquad
\beta_{j}=\frac{3}{5}(\pm\log\kappa N+\tilde{\beta}_{j}^{\mp})
\end{equation}
This was checked numerically out to the system size $N=16$. 

In the $x=\frac{5}{3i}(u-\frac{3\pi}{10})$ variable, the zeros of the 1-strings are
located at 
\begin{equation}
x_{0}^{\pm}=\frac{5\alpha}{3}=\pm\log\kappa N+\tilde{\alpha}^{\mp}
\end{equation}
and the zeros of the short 2-strings are located at 
\begin{equation}
(x_{j}^{\pm}+\frac{i\pi}{6},x_{j}^{\pm}-\frac{i\pi}{6})\quad;
\qquad x_{j}^{\pm}=\pm\log\kappa N+\tilde{\beta}_{j}^{\mp}
\end{equation}

We need to convert the lattice functional equations into TBA integral equations, solve by Fourier transforms and then take the continuum scaling limit. 
To do so we need ANZC functions that are free of zeros and poles in an open strip containing 
$\Im x\in [-\frac{\pi}{3},\frac{\pi}{3}]$. The function which removes
the single zero introduced by a 1-string is 
\begin{equation}
\sigma_{0}(x)=\tanh\frac{3x}{4}
\end{equation}
while the one which removes the two zeros of a short 2-string is
\begin{equation}
\sigma_{1}(x)=\frac{\cosh x-\sin\frac{\pi}{3}}{\cosh x+\sin\frac{\pi}{3}}
\end{equation}
 These functions satisfy~\cite{BLZ}
\begin{equation}
\sigma_{0}(x-\frac{i\pi}{3})\sigma_{0}(x+\frac{i\pi}{3})=1,\qquad
\sigma_{1}(x-\frac{i\pi}{3})\sigma_{1}(x+\frac{i\pi}{3})=\sigma_{1}(x)
\label{BLZfunct}
\end{equation}
We therefore parametrize the normalized transfer matrix eigenvalue as 
\begin{equation}
t(x)=f(x)\prod_{\pm}\sigma_{0}(x-x_{0}^{\pm})\prod_{j=1}^{M}\sigma_{1}(x-x_{j}^{\pm})\,l(x)
\end{equation}
which ensures that $l(x)$ is ANZC in the analyticity strip $\Im x\in (-\frac{5\pi}{6},\frac{5\pi}{6})$.

After removing the zeros, the functional equation takes the form
\begin{equation}
\prod_{\pm}\sigma_{0}(x-x_{0}^{\pm})\frac{t(x-i\frac{\pi}{3})t(x+i\frac{\pi}{3})}{t(x)}
=\frac{l(x-i\frac{\pi}{3})l(x+i\frac{\pi}{3})}{l(x)}=\prod_{\pm}\sigma_{0}(x-x_{0}^{\pm})
\Big(1+{1\over t(x)}\Big)
\end{equation}
Notice that the combination $\prod_{\pm}\sigma_{0}(x-x_{0}^{\pm})/t(x)$
is regular at $x=x_{0}^{\pm}.$ Taking the logarithm and using Fourier transforms again we find
\begin{equation}
\log l(x)=-\varphi\star\log\Big[\prod_{\pm}\sigma_{0}(x-x_{0}^{\pm})\Big(1+{1\over t(x)}\Big)\Big]
\end{equation}
Restoring $t(x)$ gives
\begin{equation}
\log t(x)\!=\!\log f(x)
+\sum_{\pm}\log\sigma_{0}(x-x_{0}^{\pm})+\sum_{j,\pm}
\log\sigma_{1}(x-x_{j}^{\pm})-\varphi\star\log\Big[\prod_{\pm}\sigma_{0}(x-x_{0}^{\pm})
\Big(1+{1\over t(x)}\Big)\Big]\label{eq:perexTBA}
\end{equation}
The parameters of the excited state $x_{i}=\{x_{0}^{\pm},x_{j}^{\pm}\}$
are determined self-consistently from the fact that they are zeros
of the transfer matrix: 
\begin{equation}
t(x)\Big\vert_{x=x_{i}\pm\frac{i\pi}{3}}=-1
\end{equation}
In the scaling limit, we focus on the two scaling domains at $\pm\log\kappa N$
by introducing 
\begin{equation}
e^{\epsilon^{\mp}(x)}=\lim_{N\to\infty}\sigma_{0}(x\pm\log\kappa N-x_{0}^{\pm})^{-1}t(x+\log\kappa N)
\end{equation}
It satisfies the equation
\begin{equation}
\epsilon^{\mp}(x)=e^{\mp x}+\sum_{j}\log\sigma_{1}(x-\tilde{\beta}_{j}^{\mp})-
\varphi\star\log(\sigma_{0}(x-\tilde{\alpha}^{\mp})+e^{-\epsilon^{\mp}(x)})
\end{equation}
The location of the zeros $\tilde{\alpha}^{\pm}$ and $\tilde{\beta}_{j}^{\pm}$
are self-consistently determined from the following equations
\begin{equation}
e^{\epsilon^{\pm}(x)}\sigma_{0}(x)\Big\vert_{x=\gamma^{\pm}\pm\frac{i\pi}{3}}=-1
\end{equation}
where $\gamma^{\pm}$ is either $\tilde{\alpha}^{\pm}$ or $\tilde{\beta}_{j}^{\pm}.$

The contribution of the roots to the energy can be calculated as before
\begin{equation}
E^{\pm}=\sum_{j}e^{\pm\tilde{\beta}_{j}^{\pm}}-\int_{-\infty}^{+\infty}
\frac{dx}{2\pi}e^{\mp x}\log(1+e^{-\epsilon^{\mp}(x)})
\end{equation}

\subsubsection{{Periodic case with} a seam}

In this section we point out the differences, in the case of a seam, compared to the periodic case. 
The normalization used to bring the fusion equation into the universal form 
\begin{equation}
t(u)t(u+\lambda)=1+t(u+3\lambda)
\end{equation}
introduces order $N-1$ zeros at $-\frac{\pi}{5},\frac{4\pi}{5}$
and order $N-1$ poles at $\frac{\pi}{5},\frac{2\pi}{5}$. Additionally, 
a single zero is introduced at $-\frac{\pi}{5}-\xi\equiv\frac{4\pi}{5}-\xi$, and
a short 2-string at $\frac{\pi}{5}-\xi,\frac{2\pi}{5}-\xi$ . To be able to describe the flow, we will
take $\xi$ to be pure imaginary. The bulk and seam dependent non-universal
functions can be factored out as in (\ref{eq:fgl}) and we seek an integral equation for $l$. The functions $f(u)$ and
$g(u)$, which remove the order $N$/order $1$ zeros and poles
are found to be 
\begin{equation}
f(u)=f_1(u)^{N-1},\qquad g(u)=f_1(u+\xi)
\end{equation}
In addition to (\ref{eq:ff1}) and (\ref{eq:gg1}), they also satisfy
\begin{equation}
f(u+\tfrac{\pi}{5})f(u-\tfrac{\pi}{5})=f(u),\qquad 
g(u+\tfrac{\pi}{5})g(u-\tfrac{\pi}{5})=g(u)
\end{equation}

\paragraph{Vacuum state.}
For the ground state without 1-strings, 
following the previous derivation (\ref{eq:tt/t}) based on the parametrization
(\ref{eq:fgl}), we obtain the equation
\begin{equation}
\log t(x)=\log f(x)+\log g(x)-\varphi\star\log(1+t^{-1}(x))
\end{equation}
This is the ground-state TBA on the lattice with a seam. 

In the continuum scaling limit, the contributions come from the scaling regions in the upper half-plane around $x\sim\log\kappa N$
and in the lower half-plane around $x\sim-\log\kappa N$. If we do not scale the parameter $\xi$ with $N$, it disappears from the equations in the scaling limit.
So we need to scale it by $\pm i\frac{3}{5}\log N$ to bring the order 1 zeros into the scaling region. 
It then only appears in the equations for $\epsilon^{\pm}$.
Let us consider the scaling region in the upper half-plane with 
$\xi=\frac{3}{5i}(\tilde{\xi}+\log\kappa N$) 
and center the new functions around $x\pm\log\kappa N$ as 
$e^{\epsilon^{\mp}(x)}=\lim_{N\to\infty}t(x\pm\log\kappa N)$.
Taking the continuum scaling limit $N\to\infty$ on the source term, we obtain 
the massless ground-state TBA equations in the presence of a seam
\begin{equation}
\epsilon^{\pm}(x)=e^{\pm x}+\log g(x-\tilde{\xi})-\varphi\star\log(1+e^{-\epsilon^{\pm}(x)})
\end{equation}

It is enlightening to compare this result to the massless limit of
the defect TBA equation \cite{BS}. We find that the function $g$
is related to the defect transmission factor $T_{-}(x,b)$ as 
\begin{equation}
g(x-\tilde{\xi})=T_{-}(\frac{i\pi}{2}+x,b),\qquad b=3+\frac{6i\tilde{\xi}}{\pi}
\end{equation}
In calculating the energy of the ground-state we see that $g$ only contributes to the seam energy and so formulas (\ref{eq:E-}) and (\ref{eq:E+})
for $E^{\pm}$ still hold. The extension of this analysis for excited
states is straightforward by including the source term $\log g(x-\tilde{x})$
into (\ref{eq:perexTBA}).

\subsubsection{Boundary case}

In this section, we consider the ground state and, specifically, the ground-state in the two
sectors labeled as $s=1,2$. In this case, the boundary normalization (\ref{eq:bdrynorm})
introduces order $2N$ zeros at $-\frac{\pi}{5},\frac{4\pi}{5}$ and order $2N$ 
poles at $\frac{\pi}{5},\frac{2\pi}{5}$. These are removed by
\begin{equation}
f(u)=f_{1}(u)^{2N}
\end{equation}
The boundary normalization (\ref{eq:bdrynorm}) introduces a double
zero at $u=\frac{\lambda}{2}$, poles at $u=-\frac{\lambda}{2}+\pi=
\frac{7\pi}{10}$
and $u=\frac{3\lambda}{2}-\pi=-\frac{\pi}{10}$. Observe
that the argument of the order $1$ normalization factor in (\ref{eq:bdrynorm})
  is $2u$,
thus its periodicity is $\frac{\pi}{2}$, which introduces poles at
$\frac{\pi}{5}$ and $\frac{2\pi}{5}$. In addition, there is a real
2-string with zeros occurring at $\frac{\pi}{10}$ and $\frac{\pi}{2}$.
The factor that satisfies
\begin{equation}
g_{1}(u+\tfrac{\pi}{5})g_{1}(u-\tfrac{\pi}{5})=g_{1}(u)
\end{equation}
to remove these zeros and poles is 
\begin{equation}
g_{1}(u)=-\frac{\tan^{2}(\frac{5u}{6}-\frac{\pi}{4})\tan(\frac{5u}{6}-
\frac{\pi}{12})\tan(\frac{5u}{6}-\frac{5\pi}{12})}{\tan(\frac{5u}{6}-
\frac{\pi}{6})\tan(\frac{5u}{6}-\frac{\pi}{3})}
\end{equation}
In the $x$ variable it takes the form 
\begin{equation}
g_{1}(x)=-\frac{\tan^{2}(\frac{ix}{2})\tan(\frac{ix}{2}+\frac{\pi}{6})
\tan(\frac{ix}{2}-\frac{\pi}{6})}{\tan(\frac{ix}{2}-\frac{\pi}{12})
\tan(\frac{ix}{2}+\frac{\pi}{12})}
\end{equation}
It appears in the lattice boundary TBA equation as 
\begin{equation}
\log t(x)=\log f(x)+\log g_{1}(x)-\varphi\star\log(1+t^{-1}(x))
\end{equation}
but does not contribute explicitly to the energy. In the boundary case, 
the lower and upper half-planes contribute equally due to complex conjugation symmetry in $u$.
By scaling the variables to the scaling around $\log2\kappa N$, the boundary contribution disappears from the TBA equations.

As the $s=2$ boundary is obtained by acting with a seam on the $s=1$ boundary, we can write
\begin{equation}
g_{2}(u)=g_{1}(u)f_1(u+\xi)f_1(u-\xi)
\end{equation}
Consequently, the lattice TBA equation is the same as for the $s=1$ boundary
except that the $g_{1}$ is replaced by $g_{2}$. Particularly interesting
is the scaling limit $x\mapsto x+\log2\kappa N$ if we also scale the
boundary parameter $\xi=\frac{3}{5i}(
\tilde{\xi}+\log2\kappa N)$ as we did in the defect case.
The resulting critical boundary TBA equation becomes
\begin{equation}
\epsilon^{\pm}(x)=e^{\pm x}+\log\frac{g_{2}(x-\tilde{\xi})}{g_{1}(x)}-
\varphi\star\log(1+e^{-\epsilon^{\pm}(x)})
\end{equation}
The term $\frac{g_{2}(x-\tilde{\xi})}{g_{1}(x)}$ coincides with the
scaling limit of the product of the reflection factors 
$\log R_{0}(\frac{i\pi}{2}-x)R_{1}(\frac{i\pi}{2}+x,b)$
\cite{DPTW}. The excited states can be described similarly. We will
spell out the details in the off-critical case, from which the critical
case can be easily obtained as a special limit.

\section{Massive TBA Equations}

In this section we solve, following the methods of \cite{PearN98}, the functional relations
\begin{equation}
t(u,q)t(u+\lambda,q)=1+t(u+3\lambda,q)
\label{massiveT}
\end{equation}
for the eigenvalues of the off-critical transfer matrix  to derive massive TBA equations in the continuum scaling limit. We regard the elliptic nome $q$, with $0<q<1$, as fixed and often suppress the dependence on this nome.  
The transfer matrix eigenvalues are actually doubly-periodic in the complex $u$ plane
\begin{equation}
t(u,q)=t(u+\pi,q),\qquad t(u+i\pi\epsilon,q)=t(u,q),\qquad q=e^{-\pi\epsilon}
\end{equation}
This is due to the quasi-periodicity of the elliptic Boltzmann weights and the normalization factor
\begin{equation}
\Big(\frac{s(\lambda)s(u+2\lambda)}{s(u+\lambda)s(u+3\lambda)}\Big)^{N}
\end{equation}
used to bring the equations into universal form. 
This normalization introduces order $N$ zeros and poles which
need to be removed by appropriately chosen functions.

Using periodicity and shifting $u$, we rewrite the functional equation as
\begin{equation}
t(u+\tfrac{\pi}{5},q)t(u-\tfrac{\pi}{5},q)=1+t(u,q)
\end{equation}
To remove the order $N$ zeros at $-\frac{\pi}{5},\frac{4\pi}{5}$
and poles at $\frac{\pi}{5},\frac{2\pi}{5}$, we need a function $f(u,q)$ which satisfies
\begin{equation}
f(u+\tfrac{\pi}{5},q)f(u-\tfrac{\pi}{5},q)=f(u,q),\qquad f(u+i\pi\epsilon,q)=
f(u,q),\quad f(u,q)=f_1(u,q)^N
\label{neweq:funcrelf}
\end{equation}
The required solution for $f_1(u,q)$, 
compatible with quasi-periodicity is
\begin{equation}
f_1(u,q)=-\frac{\vartheta_{1}(\frac{5u}{6}+\frac{\pi}{6},p)\vartheta_{1}(\frac{5u}{6}+\frac{2\pi}{6},p)}{\vartheta_{2}(\frac{5u}{6}+\frac{\pi}{6},p)\vartheta_{2}(\frac{5u}{6}+\frac{2\pi}{6},p)}=-\frac{\vartheta_{1}(\frac{5u}{6}+\frac{\pi}{6},p)\vartheta_{1}(\frac{5u}{6}+\frac{2\pi}{6},p)}{\vartheta_{1}(\frac{5u}{6}+\frac{4\pi}{6},p)\vartheta_{1}(\frac{5u}{6}+\frac{5\pi}{6},p)}
\label{fuq}
\end{equation}
where the periodicity, $u\equiv u+i\pi\epsilon$, requires 
\begin{equation}
p=q^{\frac{5}{6}}
\end{equation}

If we define 
\begin{equation}
t(u,q)=f(u,q)l(u,q)
\end{equation}
then the resulting function $l(u)$ is analytic and nonzero
in the required domain. Introducing the variable $x$ through
\begin{equation}
u=\frac{3\pi}{10}+\frac{3ix}{5}
\end{equation}
the functional equation becomes 
\begin{equation}
t(x-\tfrac{\pi i}{3},q)t(x+\tfrac{\pi i}{3},q)=1+t(x,q)\label{eq:funcreloff}
\end{equation}
The periodicity rectangle in the variable $x$ is $\Re x\in(-\frac{5\pi\epsilon}{6},\frac{5\pi\epsilon}{6})$
and $\Im x\in(-\frac{5\pi}{6},\frac{5\pi}{6})$. To solve
the equation by Fourier series, we need functions analytic
and nonzero in an open rectangle with the same imaginary period but containing the interval $\Im x\in(-\frac{\pi}{3},\frac{\pi}{3})$. 
This analyticity rectangle is the analogue of the analyticity strip.

\subsection{Periodic vacuum state}

\subsubsection{TBA equation}

We divide (\ref{eq:funcreloff}) by $t(x,q)$ and use the functional relation
(\ref{neweq:funcrelf}) to write 
\begin{equation}
\frac{t(x-i\frac{\pi}{3},q)t(x+i\frac{\pi}{3},q)}{t(x,q)}=\frac{l(x-i\frac{\pi}{3},q)l(x+i\frac{\pi}{3},q)}{l(x,q)}=1+t^{-1}(x,q)
\end{equation}
After taking the logarithm (both sides are ANZ in the analyticity rectangle) we solve
this equation in Fourier space. The functions are periodic in $x$ with period $V=\frac{5\pi\epsilon}{3}$. So, by Fourier inversion, such functions satisfy
\begin{equation}
h(x)=\sum_{k=-\infty}^\infty e^{i\omega kx}h_{k},\qquad h_{k}=\frac{1}{V}\int_{-\frac{V}{2}}^{\frac{V}{2}}\, h(x)e^{-i\omega kx}dx,\qquad\omega=\frac{2\pi}{V}
\end{equation}
Solving the functional equation for $\log l(x)$ gives 
\begin{equation}
\log l_{k}=\frac{\log(1+t^{-1})_{k}}{e^{\frac{\pi}{3}\omega k}+e^{-\frac{\pi}{3}\omega k}-1}
\end{equation}
or in real space 
\begin{equation}
\log l(x,q)=-\varphi_{\epsilon}\star\log(1+t^{-1}(x,q)):=-\frac{1}{V}\int_{-\frac{V}{2}}^{\frac{V}{2}}dy\,\varphi_{\epsilon}(x-y,q)\log(1+t^{-1}(y,q))
\end{equation}

\def\q{\tilde{q}}
The kernel 
\begin{equation}
\varphi_{\epsilon}(x,q)=\sum_{k=-\infty}^\infty \frac{e^{i\omega kx}}{1-e^{\frac{\pi}{3}\omega k}-e^{-\frac{\pi}{3}\omega k}}
\end{equation}
must be doubly periodic, so we write it in terms
of elliptic functions. This is achieved by writing
\begin{equation}
\varphi_{\epsilon}(x,q)=-\sum_{k=-\infty}^\infty e^{i\omega kx}\,\frac{e^{\frac{\pi}{6}\omega k}+e^{-\frac{\pi}{6}\omega k}}{e^{\frac{\pi}{2}\omega k}+e^{-\frac{\pi}{2}\omega k}}=-\sum_{k=-\infty}^\infty \frac{e^{i\omega k(x+i\frac{\pi}{6})}+e^{i\omega k(x-i\frac{\pi}{6})}}{\tilde{q}^{k}+\tilde{q}^{-k}},\qquad\tilde{q}=e^{-\frac{\pi}{2}\omega}
\end{equation}
Using formula (3) in (8.146) of \cite{GR}
\begin{equation}
\dn u=\dn(u,\q)
=\frac{\pi}{K} \sum_{k=-\infty}^\infty \frac{e^{\frac{ik\pi u}K}}{\q^k+\q^{-k}},\qquad \big|\Im\big(\frac{\pi u}{K}\big)\big| < {\pi\omega\over 2}
\end{equation}
where $K=K(\q)$ is the complete elliptic integral of nome $\q$ and $K'=K(\q')$ is the complete elliptic integral with conjugate nome $\q'$
\begin{equation}
\q=e^{-\frac{\pi K'}{K}}=e^{-\frac \pi2 \omega}=e^{-\frac{3\pi}{5\epsilon}},\qquad 
\q'=e^{-\frac{\pi K}{K'}}=e^{-\frac{2\pi}\omega}=e^{-\frac{5\pi\epsilon}{3}}=q^{\frac{5}{3}}=p^2
\end{equation}
It follows that
\begin{equation}
\varphi_{\epsilon}(x,q)=-\frac K\pi\Big[ \dn \frac{\omega K}{\pi}\big(x+\frac{\pi i}{6}\big)+\dn \frac{\omega K}{\pi}\big(x-\frac{\pi i}{6}\big)\Big]
\end{equation}
But following \cite{PearN98}
\begin{equation}
\dn \frac{2K' x}{\pi}=\frac{\pi}{2K'} \frac{\vartheta_2(0,\q')\vartheta_3(0,\q')\vartheta_3(ix,\q')}{\vartheta_2(ix,\q')}
\end{equation}
Using further that $\omega=\frac{2K'}{K}$, we arrive at the useful form 
\begin{equation}
\varphi_{\epsilon}(x,q)=-\frac{1}{\omega}\,\vartheta_{2}(0,\tilde{q}')\vartheta_{3}(0,\tilde{q}')\Big[ \frac{\vartheta_{3}(ix+\frac{\pi}{6},\tilde{q}')}{\vartheta_{2}(ix+\frac{\pi}{6},\tilde{q}')}+\frac{\vartheta_{3}(ix-\frac{\pi}{6},\tilde{q}')}{\vartheta_{2}(ix-\frac{\pi}{6},\tilde{q}')}\Big],
\quad\tilde{q}'=p^{2}
\end{equation}
In the critical limit, when $V=\frac{5\pi\epsilon}{3}\to \infty$ and $\q'\to 0$, the kernel simplifies to
\begin{align}
\varphi(x)&
=\lim_{\epsilon,V\to\infty} \frac{2\pi}{V}
\,\varphi_\epsilon(x,q)=\lim_{\epsilon\to\infty} \frac{6}{5\epsilon}\,
\varphi_\epsilon(x,q)\nonumber\\
&=-\big[{\sech(x+\tfrac{\pi i}{6})}
+{\sech(x-\tfrac{\pi i}{6})}\big]=-
\,\frac{4\sqrt{3}\cosh x}{1+2\cosh 2x}
\end{align}
which, as we have already seen, is related to the logarithmic derivative of the Lee-Yang $S$-matrix.

The final lattice TBA equation is obtained by restoring $t(x,q)$
\begin{equation}
\log t(x,q)=\log f(x,q)-\varphi_{\epsilon}\star\log(1+t^{-1}(x,q))
\end{equation}
In the continuum scaling limit, the interesting domain is
again the scaling region $x\mapsto x+\log\kappa N$ with $\kappa=2\sqrt{3}$.
To obtain a finite limit, we take the massive scaling limit
\begin{equation}
\lim_{N\to\infty,\,a\to0} Na=L,\qquad m=\lim_{t\to 0,\,a\to 0} \frac{4\sqrt{3}\,t^\nu}{a}, 
\qquad pN=q^{\frac{5}{6}}N=\frac{mL}{4\sqrt{3}}=\mbox{fixed}
\label{massivelimit}
\end{equation}
where $a$ is the lattice spacing, $t=q^2$ is the deviation from the critical temperature and $\nu=\tfrac{5}{12}$ is the correlation length exponent. As expected~\cite{PearN98}, the temperature only enters in the combination $t^\nu$. 
It then follows that    
\begin{equation}
\lim_{N\to\infty}\log f(x+\log\kappa N,q)
=e^{-x}+(\kappa Np)^{2}e^{x}+O(p^{4}N^{3})
\end{equation}
After shifting the $x$ variable, the lattice calculation yields the standard massive TBA
\begin{equation}
\log t(x)=mL\cosh x-\varphi\star\log(1+t^{-1}(x))
\end{equation}

Let us emphasize that taking $p\sim 1/N$ actually means that
the scaling regions $x\sim\log N$ and $x\sim-\log N$ are the same
by periodicity of the functions. If we had centered the functions
around $x\mapsto x+\frac{5}{6}\pi\epsilon$ then the small $p$ expansion
directly gives 
\begin{equation}
f_1(x+\tfrac{5}{6}\pi\epsilon,q)
=\frac{\vartheta_{4}(\frac{i}{2}x+\frac{5\pi}{12},p)\vartheta_{4}(\frac{i}{2}x+\frac{7\pi}{12},p)}{\vartheta_{4}(\frac{i}{2}x-\frac{\pi}{12},p)\vartheta_{4}(\frac{i}{2}x+\frac{\pi}{12},p)}=1+2\kappa p\cosh x+O(p^{2})
\end{equation}
where we used the quasi-periodicity relation
\begin{equation}
\vartheta_{1}(u-\frac{i}{2}\log p,p)=iq^{-\frac{1}{4}}e^{-iu}\vartheta_{4}(u,p)
\end{equation}
Thus defining 
\begin{equation}
\lim_{N\to\infty}\log t(x+\tfrac{V}{2},q)=\tilde{\epsilon}(x)
\end{equation}
 the TBA equation in the massive scaling limit is 
\begin{equation}
\tilde{\epsilon}(x)=mL\cosh x-\varphi\star\log(1+e^{-\tilde{\epsilon}})(x)
\end{equation}
This is the ground-state TBA equation of \cite{Zam90}  obtained directly from 
minimizing the Euclidean partition function.

\subsubsection{Energy formula}

The vacuum energy $E_{0}$ in the massive scaling limit can be obtained from the finite-size corrections on the lattice which, since there is no boundary contribution, take the form~\cite{PearN98}
\begin{equation}
-\log t(u,q)=NE_\text{\tiny bulk}(u)+\frac{1}{N}\sin\frac{5u}{3}E_{0}+\dots
=NE_\text{\tiny bulk}(u)+\frac{1}{N}\,E_{0}\cosh x+\dots
\end{equation}
We use  the TBA equation
\begin{equation}
-\log t(x,q)=-
\log f(x,q)+\frac{1}{V}\int_{-\frac{V}{2}}^{\frac{V}{2}}dy\,\varphi_{\epsilon}(x-y,q)\log(1+t^{-1}(y,q))
\end{equation}
Since the integration is over the imaginary period, we can shift the integration variable $y$ by $\pm \tfrac{1}{2} V$ without changing the terminals 
\begin{equation}
-\log t(x,q)=-\log f(x,q)+\frac{1}{V}
\int_{-\tfrac{V}{2}}^{\frac{V}{2}}dy\,\varphi_{\epsilon}(x-y\mp\tfrac{V}{2},q)\log(1+t^{-1}(y\pm \tfrac{V}{2},q))
\end{equation}
Next we observe that
\begin{equation}
\lim_{q\to0}\frac{1}{V}\varphi_{\epsilon}(x\pm\tfrac{V}{2},q)=-\frac{1}{N}\frac{m}{2\pi}\cosh x
\end{equation}
So, taking half the sum of the two shifts, we find in the massive scaling limit (\ref{massivelimit})
\begin{equation}
E_\text{\tiny bulk}(x)=-\log f_1(x,0)
\end{equation}
and 
\begin{equation}
E_{0}=-m\int_{-\infty}^{\infty}\frac{dy}{2\pi}\cosh y\log(1+e^{-\tilde{\epsilon}(y)})
\end{equation}
where we used that $\epsilon(x)=\epsilon(-x)$ and $\frac{1}{2}[\cosh(x-y)+\cosh(x+y)]=\cosh x\cosh y$.

\subsection{Periodic excited states}

\subsubsection{TBA equation}

For excited states we assume, as supported by numerics, the existence
of real 1-strings and short 2-strings. 
The relevant functions to remove
these zeros with the required periodicity are respectively
\begin{equation}
\sigma_{0}(x,q)=i\,\frac{\vartheta_{1}(\frac{3ix}{4},q^{\frac{5}{4}})}{\vartheta_{2}(\frac{3ix}{4},q^{\frac{5}{4}})}
\end{equation}
\begin{equation}
\sigma_{1}(x,y,q)=-\frac{\vartheta_{1}(\frac{ix}{2}-\frac{y}{2},q^{\frac{5}{6}})\vartheta_{1}(\frac{ix}{2}+\frac{y}{2},q^{\frac{5}{6}})}{\vartheta_{2}(\frac{ix}{2}-\frac{y}{2},q^{\frac{5}{6}})\vartheta_{2}(\frac{ix}{2}+\frac{y}{2},q^{\frac{5}{6}})}
\end{equation}
The elliptic nomes are fixed to maintain the same imaginary period as the transfer matrix eigenvalues. 
We note that $\sigma_{0}(x,q)$ has a single zero at $x=0$ within the period
rectangle, whereas $\sigma_{1}(x,y,q)$ has two zeros at $x=\pm i y$.
If a short 2-string occurs at $x_j\pm i\tfrac{\pi}{6}$, then it can be removed using $\sigma_1(x,\tfrac{\pi}{6},q)$. 
If a short 2-string occurs at $x_j\pm iy_j$, with $y_j>\tfrac{\pi}{6}$, then following \cite{BLZ} it can be 
removed by the function
\begin{equation}
\tilde{\sigma}_{1}(x_j,y_j,q)=\sigma_{1}(x_j,y_j,q)\sigma_1(x_j,y_j+\tfrac{\pi}{3},q)
\end{equation}
The functions $\sigma_0(x,q)$, $\sigma_1(x,\frac{\pi}{6},q)$ and $\tilde{\sigma}_1(x,y,q)$ satisfy the functional relations (\ref{BLZfunct}). 
Now suppose that we have 1-strings at $x_{0}^{\pm}$ and short 
2-strings at $(x_{j}^{\pm},y_{j}^{\pm})$ where the superscripts $\pm$ indicate that they lie in the upper and lower half-planes respectively. Then we can write   
$t(x,q)$ as 
\begin{equation}
t(x,q)=f(x,q)
\prod_{\{\pm\}}\sigma_{0}(x-x_{0}^{\pm},q)\prod_{j}\tilde{\sigma}_{1}(x-x_{j}^{\pm},y_{j}^{\pm},q)l(x,q)
\end{equation}
A calculation analogous to the massless case leads to 
\begin{align}
\log t(x)&=\log f(x)+\sum_{\pm}\log
(\sigma_{0}(x-x_{0}^{\pm},q))
+\sum_{j,\pm}\log(\tilde{\sigma}_{1}(x-x_{j}^{\pm},y_{j}^{\pm},q))\nonumber\\
&\qquad-\varphi_{\epsilon}
\star\log\Big(\prod_{\pm}\sigma_{0}(x-x_{0}^{\pm},q)(1+t^{-1}(x))\Big)
\end{align}

The TBA equations in the continuum limit are obtained by centering
the functions around $\frac{V}{2}$. In the continuum scaling limit $q\to 0$ and $V\to\infty$
but the roots also scale with $\log N$ and remain around $\frac{V}{2}$:
$x_{j}=\tilde{x}_{j}+\frac{V}{2}$ ($\tilde{x}_{j}$ remaining finite).
Let us define the scaling functions
\begin{equation}
e^{\tilde{\epsilon}(x)}=\lim_{V\to\infty}\sigma_{0}(x\pm\tfrac{V}{2}-x_{0}^{\pm})^{-1}t(x+\tfrac{V}{2})
\end{equation}
which satisfies the equation
\begin{equation}
\tilde{\epsilon}(x)=mL\cosh x+\sum_{j}\log\tilde{\sigma}_{1}(x-\tilde{x}_{j},{y}_{j})-\varphi\star
\log(\sigma_{0}(x-\tilde{x}_{0})+e^{-\tilde{\epsilon}(x)})
\end{equation}
where $\tilde{\sigma}_1(x,y)=\lim_{q\to 0}\tilde\sigma_1(x,y,q)$. 
From the $Y$-system, the equations determining the roots are
\begin{equation}
e^{\tilde{\epsilon}(x)}\sigma_{0}(x)\vert_{x=x_j+iy_j\pm\frac{i\pi}{3}}=-1
\end{equation}
These equations coincide with \cite{BLZ}.

We can compare our results with the TBA equation obtained by analytic
continuation in \cite{Dorey}. For large volumes, $mL\gg1$, we find that 
the two 1-strings disappear and the TBA equation coincides with \cite{Dorey}
\begin{equation}
\tilde{\epsilon}(x)=mL\cosh x+\sum_{j}\log\frac{S(x-\theta_{j})}{S(x-\theta_{j}^{\star})}-\varphi\star\log(1+e^{-\tilde{\epsilon}(x)})
\label{massiveperiodicTBA}
\end{equation}
where the rapidities are
\begin{equation}
\theta_{j}=\tilde{x}_{j}+i(\tfrac{\pi}{3}-{y}_{j}),\qquad\theta_{j}^{\star}
=\tilde{x}_{j}-i(\tfrac{\pi}{3}-{y}_{j})
\end{equation}
and we used (\ref{Smatrix}). 

\subsubsection{Energy formula}

The energy of the massive Lee-Yang theory is obtained from the relation
\begin{equation}
-\log t(x,q)=-\log f(x,q)-\sum_{j}
\log\tilde{\sigma}_{1}(x-x_{j},y_{j},q)-\log l(x,q)
\end{equation}
where, setting $x_{j}=\frac{V}{2}+\tilde{x}_{j}$, 
\begin{equation}
\sigma_{1}(x+\tfrac{V}{2},y,q)=-\frac{\vartheta_{1}(\frac{ix}{2}-\frac{y}{2},p)\vartheta_{1}(\frac{ix}{2}+\frac{y}{2},p)}{\vartheta_{2}(\frac{ix}{2}-\frac{y}{2},p)\vartheta_{2}(\frac{ix}{2}+\frac{y}{2},p)}
\bigg\vert_{x\mapsto x+\frac{V}{2}}
=\frac{\vartheta_{4}(\frac{ix}{2}-\frac{y}{2},p)\vartheta_{4}(\frac{ix}{2}+\frac{y}{2},p)}
{\vartheta_{3}(\frac{ix}{2}-\frac{y}{2},p)\vartheta_{3}(\frac{ix}{2}+\frac{y}{2},p)}
\end{equation}
Taking the massive scaling limit (\ref{massivelimit}) with $q\to 0$, $N\to\infty$,  gives 
\begin{equation}
\log\tilde{\sigma}_{1}(x+\tfrac{V}{2},y,q)=-8\sqrt{3} p\cosh x \cos (y+\frac {\pi}{6})+O(p^3)
=-\frac{2mL}{N}\cosh x\cos (y+\frac {\pi}{6})+O(N^{-3})
\end{equation}
Repeating a similar calculation as for the ground-state, we obtain \begin{align}
-\log t(x,q) & =  -\log f(x,0)
-m\int_{-\infty}^{\infty}\frac{dy}{2\pi}\cosh(x-y)\log(1+e^{-\tilde{\epsilon}(y)})\nonumber \\
 & +\frac{mL}{N}\sum_{j}\left[\cosh(x-\tilde{x}_{j}
 -iy_{j}-\tfrac{i\pi}{6})+\cosh(x-\tilde{x}_{j}+iy_{j}
 +\tfrac{i\pi}{6})\right]
\end{align}
For the isotropic point $x=0$, we obtain the energy formula of \cite{Dorey}
\begin{equation}
E(L)=-im\sum_{j}(\sinh\theta_{j}-\sinh\theta_{j}^{*})-m\int_{-\infty}^{\infty}\frac{d\theta}{2\pi}\,\cosh\theta\,\log(1+e^{-\tilde{\epsilon}(\theta)})
\end{equation}

\subsection{Defect TBA equations}

In this section we derive the off-critical TBA equation on the circle with a seam. As the considerations are very similar to the periodic case we 
emphasize only the differences.  We also follow the analysis of the critical seam case. 

The functional equation to solve is the same as in the periodic case
(\ref{massiveT}).  The main difference is the analytic structure of $\vec T^s(u)$ (\ref{seamnorm}) 
arising from the normalization factor. 
The ``bulk'' zeros and poles are now only of order  $N-1$. Additionally, there is also a  
single zero at $-\frac{\pi}{5}-\xi\equiv\frac{4\pi}{5}-\xi$ and
two poles at $\frac{\pi}{5}-\xi,\frac{2\pi}{5}-\xi$. We remove these
zeros and poles by defining
\begin{equation}
t(u,q)=f(u,q)
g(u,q)l(u,q),\qquad f(u,q)=f_1(u,q)^{N-1},\quad g(u,q)=f_1(u+\xi,q)
\end{equation}
where $f_1(u,q)$ is the same as in the periodic case (\ref{fuq}).
The derivation of the vacuum lattice TBA equation with a seam is analogous to the periodic calculation, namely, we solve the 
functional equation for $l$  by Fourier transforms. The result in terms of $t(x,q)$ is 
\begin{equation}
\log t(x,q)=\log  f(x,q)+\log g(x,q)-\varphi_{\epsilon} 
\star \log (1+t^{-1}(x))
\end{equation} 

In the continuum scaling limit for fixed $\xi$, the seam dependent source term disappears since $\log g(x) \to 0$. 
However, if we scale $\xi$ as 
$\tilde \xi=\frac{V}{2}+\frac{5 i\xi}{3}$ then $g(x,q)$ will scale to the criticial $g(x)$ and the 
the massive TBA equation for the ground-state with a seam
becomes
\begin{equation}
\log t(x)=mL\cosh x +\log g(x-\tilde \xi)-\varphi\star
\log(1+t^{-1}(x))
\end{equation}
This equation is the same as obtained directly in \cite{BS} from the partition function,  with the identification
\begin{equation}
g(x-\tilde{\xi})=T_{-}(\tfrac{i\pi}{2}+x,b),\qquad b=3+\tfrac{6i\tilde{\xi}}{\pi}
\end{equation}
This result confirms the conjecture for the transmission factor in  \cite{BS}, which was obtained by 
the fusion principle. 

The excited states TBA equations are analogous to (\ref{massiveperiodicTBA}), we merely insert a 
$\log g(x)$ source term as we did for the ground state equation. The energy formula is not changed as
$g(x)$ contributes only to the defect energy.

\subsection{Boundary  TBA equations}

In this section we repeat the analysis of the previous section for the boundary case. We focus only on the extra source arising from the boundary. 

The zero and pole structure of the off-critical case is similar to the critical case. 
The main difference is the periodicity of the functions. In particular, since the normalization factor 
in (\ref{seamnorm}) has argument $2u$, it introduces boundary related $O(1)$ poles not only
on the real line, but also at half of the periodicity rectangle, i.e. at $\Im u=\tfrac{1}{2}\log q$. Thus when we 
choose the parametrization
\begin{equation}
t(u,q)=f(u,q)g_{1}(u,q)l(u,q),\quad f(u,q)=f_1(u,q)^{2N}
\end{equation}
and replace $\tanh x\mapsto \frac{\vartheta_{1}(x,p')}{\vartheta_{2}(x,p')}$ to obtain 
\begin{equation}
g{}_{1}(x,q)=-\frac{\vartheta_{1}^{2}
(\frac{ix}{2},p')\vartheta_{1}(\frac{ix}{2}
+\frac{\pi}{6},p')\vartheta_{1}(\frac{ix}{2}-\frac{\pi}{6},p')\vartheta_{2}(\frac{ix}{2}
-\frac{\pi}{12},p')\vartheta_{2}(\frac{ix}{2}+\frac{\pi}{12},p')}{\vartheta_{2}^{2}(\frac{ix}{2},p')
\vartheta_{2}(\frac{ix}{2}+\frac{\pi}{6},p')\vartheta_{2}(\frac{ix}{2}-\frac{\pi}{6},p')\vartheta_{1}
(\frac{ix}{2}-\frac{\pi}{12},p')\vartheta_{1}(\frac{ix}{2}+\frac{\pi}{12},p')}
\end{equation}
this will remove all the unwanted poles and zeros by choosing the nome $p'=q^{\frac{5}{12}}$. 
The derivation of the lattice ground-state TBA equation is analogous to the periodic case and leads to 
\begin{equation}
\log t(x,q)=\log  f(x,q) +\log g_{1}(x,q)
-\varphi_{\epsilon} \star \log (1+t^{-1}(x))
\end{equation} 

Due to the double periodicity of the boundary source term 
$g_{1}(x+\frac{V}{2},q)=g_{1}(x,q)$ the boundary 
TBA equation in the continuum scaling limit takes the form
\begin{equation}
\log t(x)=2mL\cosh x +\log g_{1}(x)-\varphi\star
\log(1+t^{-1}(x))
\end{equation}
where $\log g_{1}(x)$ is the trigonometric 
limit of $ \log g_{1}(x,q)$. This coincides with 
the boundary chemical potential $-\log (R(i \frac{\pi}{2}+x)R(i \frac{\pi}{2}-x))$ in the BTBA equation in \cite{DPTW}, 
where $R(x)$ is the reflection factor of the identity boundary condition. 

In deriving the excited state BTBA equations we have to keep in mind, that in the boundary case, strings 
appear in the $u$ variable in complex conjugates about the real line.  So, whenever we have a source term with ${\tilde x}_j$, we must also have another source term with $-{\tilde x}_j$. 

In the energy formula there is an extra half compared to the periodic case, which ensures the correct dispersion 
relation in the large volume limit. The boundary term does not contribute to the energy, only to the boundary 
energy.

\section{Discussion}

In this paper, we analysed the simplest nontrivial relativistic integrable theory, namely the Lee-Yang model,  from the lattice point of view in three different space-time geometries: on the interval, on the circle with either periodic boundary conditions or with a defect inserted. 
At its critical point, the Lee-Yang model describes the simplest non-unitary minimal model ${\cal M}(2,5)$ as the continuum scaling limit of the associated $ A_{4}$ RSOS Forrester-Baxter model with trigonometric weights. We classified all of the states of the theory by the patterns of zeros of the corresponding transfer matrix eigenvalues. Alternatively, the states are also classified by RSOS paths related to one-dimensional configurational sums and by conformal Virasoro states. The introduction of a boundary/defect allows for a boundary field $\xi$ and thus a one parameter family of perturbations. We studied the resulting  renormalization group flows, which connect different boundary/defect conditions. These flows exactly  reproduce the boundary and defect flows observed (numerically) using the boundary TCSA in \cite{DPTW} and the defect TCSA in \cite{Bajnok:2013waa}.

By considering off-critical elliptic lattice Boltzmann weights, we also analysed the 
massive integrable $\Phi_{1,3} $ perturbation of the Lee-Yang model in the three finite geometries. 
For all boundary conditions and irrespective of the geometries, both at and off-criticality, the transfer matrices satisfy the same universal~\cite{universal} $Y$ system. By extracting carefully the relevant  analytic information from the lattice in the various circumstances, we could turn the $Y$ system functional equations into TBA integral equations. 
The various TBA equations differ from each other only in the source terms. They describe exactly the finite-size scaling spectra of the Lee-Yang model in the continuum scaling limit. The derived integral equations 
confirm the conjectured excited state TBA equations for the periodic~\cite{Dorey} and boundary cases ~\cite{DPTW} and agree with the ground-state defect TBA equations confirming the transmission factor of \cite{Bajnok:2013waa}. 

The lattice description of the simplest integrable scattering theory enables the determination of the spectrum. However, this framework also establishes a solid starting point for investigating other interesting and relevant physical quantities such as vacuum expectation values and form factors, for which results from the bootstrap approaches are available \cite{Bajnok:2009hp,Bajnok:2013eaa}. Conceivably, this approach could also give insight into the calculation of correlations functions. A particularly interesting problem is the calculation~\cite{FranchiniDeLuca,BianchiniThesis} of the boundary/entanglement entropy from the lattice.
The Lee-Yang model is just the first member of the series of non-unitary minimal models ${\cal M}(m,m')$. 
It is an interesting and timely problem to extend the results of this paper to the other non-unitary minimal models.

\acknowledgments

ZB and OeD were supported by OTKA 81461 and ZB by a Lend\" ulet Grant. 
PAP is supported by the Melbourne University Research Grant Support Scheme and gratefully acknowledges support during a visit to the Asia Pacific Centre for Theoretical Physics, Pohang, South Korea.


\begin{thebibliography}{10}

\bibitem{BY} C.N. Yang, 
{\em Some exact results for the many-body problem in one dimension with repulsive delta function interaction}, 
Phys. Rev. Lett. {\bf 19} (1967) 1312.

\bibitem{YangYang} C.N. Yang and C.P. Yang, 
{\em Thermodynamics of a one-dimensional system of bosons with repulsive delta function interaction}, 
J. Math. Phys. \textbf{10} (1969) 1115.

\bibitem{Zam90} Al.B. Zamolodchikov, 
{\em Thermodynamic Bethe ansatz in relativisitic models: Scaling 3-state Potts and Lee-Yang models}, 
Nucl. Phys. \textbf{B342} (1990) 695--720

\bibitem{Zam91} Al.B. Zamolodchikov, 
{\em Thermodynamic Bethe ansatz for RSOS scattering theories}, 
Nucl. Phys. \textbf{B358} (1991) 497--523; 
{\em From tricritical Ising to critical Ising by thermodynamic Bethe ansatz}, 
Nucl. Phys. \textbf{B358} (1991) 524--546; 
{\em TBA equations for integrable perturbed $SU(2)_k\times SU(2)_\ell/SU(2)_{k+\ell}$ models}, 
Nucl. Phys. \textbf{B366}
(1991) 122--132.

\bibitem{Dorey} P. Dorey and R. Tateo, 
{\em Excited states by analytic continuation of TBA equations}, 
Nucl. Phys. \textbf{B482} (1996) 639--659.

\bibitem{LeeYang} T.D. Lee and C.N. Yang, 
{\em Statistical theory of equations of state and phase transitions. I. Theory of condensation ; II. Lattice gas and Ising model}, 
Phys. Rev. {\textbf 87} (1952) 404; 410.

\bibitem{Fisher78} M.E. Fisher, 
{\em Yang-Lee edge singularity and $\phi^3$ field theory}, 
Phys. Rev. Lett. {\bf 40} (1978) 1610.

\bibitem{Cardy85} J.L. Cardy, 
{\em Conformal invariance and the Yang-Lee edge singularity in two dimensions}, 
Phys. Rev. Lett. {\bf 54} (1985) 1354.

\bibitem{CardyMuss89} J.L. Cardy, G. Mussardo, 
\emph{$S$ matrix of the Yang-Lee edge singularity in two-dimensions}, 
Nucl. Phys. \textbf{B225} (1989) 275.

\bibitem{PearK91} P.A. Pearce and A. Kl\"umper, 
{\em Finite-size corrections and scaling dimensions of solvable lattice models: An analytic method}, 
Phys. Rev. Lett. \textbf{66} (1991) 974.

\bibitem{KlumP91} A. Kl\"umper and P.A. Pearce, 
{\em Analytic calculation of scaling dimensions: Tricritical hard squares and critical hard hexagons}, 
J. Stat.\ Phys.\ \textbf{64} (1991) 13--76.

\bibitem{KlumP92} A. Kl\"umper and P.A. Pearce, 
{\em Conformal weights of RSOS lattice models and their fusion hierarchies}, 
Physica \textbf{A183} (1992) 304.

\bibitem{BaxBook} R.J. Baxter, Exactly Solved Models in Statistical
Mechanics. Academic Press, London, 1982.

\bibitem{OPW97} D.L. O'Brien, P.A. Pearce and S.O. Warnaar, 
{\em Calculation of conformal partition funtions: Tricritical hard squares with fixed boundaries}, 
Nucl.Phys. \textbf{B501}, 773--799 (1997).

\bibitem{PearN98} P.A. Pearce and B. Nienhuis, 
{\em Scaling limit of RSOS lattice models and TBA equations}, 
Nucl.\ Phys.\ \textbf{B519} (1998) 579.

\bibitem{PCAI} P.A. Pearce, L. Chim and C. Ahn, \textit{Excited
TBA equations I: massive tricritical Ising model}, hep-th/0012223,
Nucl. Phys. \textbf{B601}, 539-568 (2001).

\bibitem{PCAII} P.A. Pearce, L. Chim and C. Ahn, 
{\em Excited TBA equations II: massless flow from tricritical to critical Ising model}, 
Nucl. Phys. \textbf{B660}, 579--606 (2003).

\bibitem{DdV} C. Destri, H.J. de Vega, 
{\em New thermodynamic Bethe ansatz equations without strings}, 
Phys.Rev.Lett. \textbf{69} (1992) 2313. 

\bibitem{DdV2} C. Destri, H.J. de Vega, 
{\em Nonlinear integral equation and excited states scaling functions in the sine-Gordon model},
Nucl.Phys. \textbf{B504} (1997) 621--664.

\bibitem{FMQR} D. Fioravanti, A. Mariottini, E. Quattrini, F. Ravanini,
\emph{Excited state Destri-De Vega equation for Sine-Gordon and restricted
Sine-Gordon models}, 
Phys.Lett. \textbf{B390} (1997) 243.

\bibitem{BLZ} V.V. Bazhanov, S.L. Lukyanov, A.B. Zamoldchikov, 
{\em Quantum field theories in finite volume: Excited state energies}, 
Nucl. Phys. \textbf{B489} (1997) 487--531.

\bibitem{LMSS} A. LeClair, G. Mussardo, H. Saleur, S. Skorik, 
{\em Boundary energy and boundary states in integrable quantum field theories},
Nucl.Phys. \textbf{B453} (1995) 581--618.

\bibitem{DPTW} P. Dorey, A.J. Pocklington, R. Tateo, G. Watts, 
{\em TBA and TCSA with boundaries and excited states},
Nucl.Phys. \textbf{B525} (1998) 641--663.

\bibitem{DRTW} P. Dorey, I. Runkel, R. Tateo and G. Watts, 
{\em g-function flow in perturbed boundary conformal field theories}, 
Nucl. Phys. {\bf B578} (2000) 85--122.

\bibitem{BS} Z. Bajnok, Zs. Simon,
{\em Solving topological defects via fusion}, 
Nucl.Phys. \textbf{B802} (2008) 307--329.

\bibitem{BPZ84} A.A. Belavin, A.M. Polyakov and A.B. Zamolodchikov, 
{\em Infinite conformal symmetry in two-dimensional quantum field theory}, 
Nucl.\ Phys.\ \textbf{B241} (1984) 333--380.

\bibitem{Huse84} D.A. Huse, 
{\em Exact exponents for infinitely many new multi critical points}, 
Phys.\ Rev.\ \textbf{B30} (1984) 3908.

\bibitem{Riggs89} H. Riggs, 
{\em Solvable lattice models with minimal and non unitary critical behaviour in two dimensions}, 
Nucl. Phys. \textbf{B326} (1989) 673--688.

\bibitem{ABF84} G.E. Andrews, R.J. Baxter and P.J. Forrester, 
{\em Eight-vertex SOS model and genderalized Rogers-Rmanujan-type identities}, 
J. Stat.\ Phys.\ \textbf{35} (1984) 193--266.

\bibitem{FB85} P.J. Forrester and R.J. Baxter, 
{\em Further exact solutions of the Eight-vertex SOS model and generalisations of the Rogers-Ramanujan identities}, 
J. Stat. Phys. \textbf{38} (1985) 435--472.

\bibitem{FPR} G. Feverati, P.A. Pearce and F. Ravanini, 
{\em Exact $\varphi_{1,3}$ boundary flows of the tricritical Ising model}, 
Nucl. Phys. {\bf B675} (2003) 469--515.

\bibitem{GR} I.S. Gradshteyn and I.M. Ryzhik, 
Tables of Integrals,Series and Products, Academic Press, 1980.

\bibitem{BEPR} D. Bianchini, E. Ercolessi, P.A. Pearce and F. Ravanini, 
{\em RSOS quantum chains associated with off-critical minimal models and $\mathbb{Z}_n$ parafermions}, 
in preparation (2014).

\bibitem{BaxP82} R.J. Baxter and P.A. Pearce, 
{\em Hard hexagons: interfacial tension and correlation length}, 
J. Phys.\ \textbf{A15} (1982) 897--910.

\bibitem{BaxP83} R.J. Baxter and P.A. Pearce, 
{\em Hard squares with diagonal attractions}, 
J. Phys.\ \textbf{A16} (1983) 2239--2255.

\bibitem{Bax80} R.J. Baxter, 
{\em Hard hexagons: exact solution}, 
J. Phys. \textbf{A13} (1980) L61.

\bibitem{BCN} H.W.J.~Bl\"ote, J.L.~Cardy, M.P. Nightingale, 
{\em Conformal invariance, the central charge, and universal finite-size amplitudes at criticality}, 
Phys. Rev. Lett. {\bf 56} (1986) 742--745.

\bibitem{Stroganov} Yu. G. Stroganov, 
{\em A new calculation method for partition functions in some lattice models}, 
Phys.\ Lett.\ \textbf{A74} (1979) 116--118.

\bibitem{Bax82Inv} R.J. Baxter, 
{\em The inversion relation method for some two-dimensional exactlyy solved models in lattice statistics}, 
J. Stat.\ Phys.\ \textbf{28} (1982) 1--41.

\bibitem{OBrienP97} D.L. O'Brien and P.A. Pearce, 
{\em Surface free energies, interfacial tensions and correlation lengths of the ABF models}, 
J. Phys. \textbf{A30} (1997) 2353--2366.

\bibitem{CMOP} C.H.O. Chui, C. Mercat, W.P. Orrick and P.A. Pearce, 
{Integrable lattice realisations of conformal twisted boundary conditions}, 
Phys. Lett. B {bf 517} (2001) 429--435.

\bibitem{BPO96} R.E. Behrend, P.A. Pearce and D.L. O'Brien, 
{\em Interaction-round-a-face models with fixed boundary conditions: The ABF fusion hierarchy}, 
J. Stat. Phys. \textbf{84} (1996) 1--48.

\bibitem{BP00} R.E. Behrend and P.A. Pearce, 
{\em Integrable and Conformal
Boundary Conditions for $s\ell(2)$ $A$-$D$-$E$ Lattice Models
and Unitary Minimal Conformal Field Theories}, 
J. Stat. Phys. {\bf 102} (2001) 577--640.

\bibitem{FNO} B.L Feigin,T. Nakanishi and H. Ooguri, 
{\em The annihilating ideals of minimal models}, 
Int. J. Mod. Phys. \textbf{ A7} (1992) 217. 

\bibitem{FodaW} O. Foda and T.A. Welsh, 
{\em On the combinatorics of Forrester-Baxter models}, 
Physical Combinatorics (Kyoto, 1999), Progress in Mathematics {\bf 191} (2000) 49--103, 
Birkhauser, Boston, MA.

\bibitem{FevPW} G. Feverati, P.A. Pearce and N.S. Witte, 
{\em Physical combinatorics and quasiparticles}, 
J. Stat. Mech. (2009) P10013.

\bibitem{BaxterCTM} R.J. Baxter, 
{\em Corner transfer matrices of the eight-vertex model I. Low temperature expansions and conjectured properties}, 
J. Stat. Phys. {\bf 15} (1976) 485--503; 
{\em Corner transfer matrices of the eight-vertex model II. The Ising model case}, 
J. Stat. Phys. {\bf 17} (1977) 1--14.

\bibitem{Fortin:2004ct} J.F. Fortin, P. Jacob, and P. Mathieu, 
{\em ${\cal SM}(2,4\kappa)$ fermionic characters and restricted jagged partitions}, 
J. Phys. A38 (2005) 1699--1710.

\bibitem{Nahm:1992sx} W. Nahm, A. Recknagel and M. Terhoeven, 
{\em Dilogarithm identities in conformal field theory}, 
Mod. Phys. Lett. {\bf A8} (1993) 1835--1848.

\bibitem{Melzer} E. Melzer, 
{\em Fermionic character sums and the corner transfer matrix}, 
Int. J. Mod. Phys. \textbf{A9} (1994) 1115.

\bibitem{Berk94} A. Berkovich, 
{\em Fermionic counting of RSOS states and Virasoro character formulas for the unitary minimal series $M(v,v+1)$: Exact results}, 
Nucl.\ Phys.\ \textbf{B431} (1994) 315--348.

\bibitem{FP02} G. Feverati and P.A. Pearce, 
{\em Critical RSOS and minimal models: fermionic paths, Virasoro algebra and fields}, 
Nucl. Phys. {\bf B663} (2203) 409--442.

\bibitem{FPW} G. Feverati, P.A. Pearce and N.S. Witte, 
{\em Physical combinatorics and quasiparticles}, 
J. Stat. Mech. (2009) P10013.

\bibitem{AffleckLudwig}
I.\ Affleck and A.W.W.\ Ludwig,
{\em Universal noninteger ``Ground-State Degeneracy'' in critical quantum systems},
Phys. Rev. Lett. {\bf 67} (1991) 161--164.

\bibitem{Bajnok:2013waa} Z.\ Bajnok, L.\ Hollo and G.\ Watts,
{\em Defect scaling Lee-Yang model from the perturbed DCFT point of view},
Nucl.\ Phys.\ B {\bf 886}, (2014) 93.

 \bibitem{universal} 
 C.H.O. Chui, C. Mercat, P.A. Pearce, 
 {\em Integrable boundaries and universal TBA functional equations}, 
 Prog. Math. Phys. {\bf 23} (2002) 391--413,
 arXiv:hep-th/0108037.

\bibitem{Bajnok:2009hp} 
  Z.~Bajnok and O.~el Deeb,
  {\em Form factors in the presence of integrable defects,}
  Nucl.\ Phys.\ B {\bf 832}, (2010) 500.
  
\bibitem{Bajnok:2013eaa} 
  Z.~Bajnok, F.~Buccheri, L.~Hollo, J.~Konczer and G.~Takacs,
  {\em Finite volume form factors in the presence of integrable defects,}
  Nucl.\ Phys.\ B {\bf 882}, (2014) 501.
  
\bibitem{FranchiniDeLuca} A. De Luca and F. Franchini, 
{\em Approaching the restricted solid-on-solid critical points through entanglement: One model for many universalities}, 
Phys. Rev. B {\bf 87} (2103) 045118.

\bibitem{BianchiniThesis} D. Bianchini, 
{\em Entanglement Entropy in Restricted Integrable Spin Chains}, M.Sc. Thesis (2013), 
University of Bologna; D. Bianchini and F. Ravanini, in preparation (2014).

\end{thebibliography}
\end{document}